\documentclass[%
 reprint,
 amsmath,amssymb,
 aps,
]{revtex4-2}

\usepackage{graphicx}% Include figure files
\usepackage{dcolumn}% Align table columns on decimal point
\usepackage{bm}% bold math
\usepackage{float}
\usepackage{tabularx}
\usepackage{setspace}
\usepackage{outlines}
\usepackage{bbm}

\usepackage{xr}
\makeatletter
\newcommand*{\addFileDependency}[1]{% argument=file name and extension
  \typeout{(#1)}
  \@addtofilelist{#1}
  \IfFileExists{#1}{}{\typeout{No file #1.}}
}
\makeatother
\newcommand*{\myexternaldocument}[1]{%
    \externaldocument{#1}%
    \addFileDependency{#1.tex}%
    \addFileDependency{#1.aux}%
}
\myexternaldocument{SI}
\usepackage{appendix}

\newcommand{\bl}{{\boldsymbol{\lambda}}}
\newcommand{\llangle}{\left\langle}
\newcommand{\rrangle}{\right\rangle}

\begin{document}

\singlespacing

\title{Learning to control non-equilibrium dynamics using local imperfect gradients}

% Force line breaks with \\
\author{Carlos Floyd}
\email{csfloyd@uchicago.edu}
\affiliation{The Chicago Center for Theoretical Chemistry, The University of Chicago, Chicago, Illinois 60637, USA
}
\affiliation{Department of Chemistry, The University of Chicago, Chicago, Illinois 60637, USA
}
\affiliation{The James Franck Institute, The University of Chicago, Chicago, Illinois 60637, USA
}
\author{Aaron R.\ Dinner}
\affiliation{The Chicago Center for Theoretical Chemistry, The University of Chicago, Chicago, Illinois 60637, USA
}
\affiliation{Department of Chemistry, The University of Chicago, Chicago, Illinois 60637, USA
}
\affiliation{The James Franck Institute, The University of Chicago, Chicago, Illinois 60637, USA
}
\author{Suriyanarayanan Vaikuntanathan}
\email{svaikunt@uchicago.edu}
\affiliation{The Chicago Center for Theoretical Chemistry, The University of Chicago, Chicago, Illinois 60637, USA
}
\affiliation{Department of Chemistry, The University of Chicago, Chicago, Illinois 60637, USA
}
\affiliation{The James Franck Institute, The University of Chicago, Chicago, Illinois 60637, USA
}

\date{\today}% It is always \today, today,

\begin{abstract}
Standard approaches to controlling dynamical systems involve biologically implausible steps such as backpropagation of errors or intermediate model-based system representations.  Recent advances in machine learning have shown that ``imperfect'' feedback of errors during training can yield test performance that is similar to using full backpropagated errors, provided that the two error signals are at least somewhat aligned.  Inspired by such methods, we introduce an iterative, spatiotemporally local protocol to learn driving forces and control non-equilibrium dynamical systems using imperfect feedback signals. We present numerical experiments and theoretical justification for several examples.  For systems in conservative force fields that are driven by external time-dependent protocols, our update rules resemble a dynamical version of contrastive divergence.  We appeal to linear response theory to establish that our imperfect update rules are locally convergent for these conservative systems.  Finally, we show that similar local update rules can also solve dynamical control problems for non-conservative systems, and we illustrate this in the non-trivial example of active nematics.  Our updates allow learning spatiotemporal activity fields that pull topological defects along desired trajectories in the active nematic fluid.  These imperfect feedback methods are information efficient and in principle biologically plausible, and they can help extend recent methods of decentralized training for physical materials into dynamical settings.  

\end{abstract}

\maketitle

\begin{twocolumngrid}
\section{Introduction}
Modern machine learning techniques have enabled unprecedented advancements in pattern recognition, decision making, generative data synthesis, and numerous other tasks \cite{lecun2015deep}.  Efficiently training models with a large number of parameters has been key for this progress, and the predominant training method involves recursively computing gradients of a global error function through automatic differentiation algorithms like backpropagation.  There is a conceptual tension, however, between this successful procedure for training artificial neural networks and our current knowledge of how synaptic weights in the brain are updated \cite{lillicrap2020backpropagation}.  Automatic differentiation of a global cost function requires knowing weights which are arbitrarily far downstream of a given unit.  This represents a top-down approach to training which is not biologically plausible given the local connectivity and plasticity mechanisms of actual neurons.  

This tension has motivated research into alternative training algorithms which avoid exact computation of a global cost function's derivatives \cite{lillicrap2020backpropagation, stern2023learning}.  In the context of a feedforward neural network architecture, recent work has shown that severe approximations to the ``correct'' gradient signal, even including a fixed random backward weight layer, can successfully be used during training if combined with accurate forward passes through the model \cite{lillicrap2016random}.  This principle of using imperfect, but easily accessible, error gradients has recently been leveraged to train real physical systems to act as machine learning models by using approximate differentiable simulations of the forward path to compute the error gradient \cite{wright2022deep}.  More broadly, the feedforward architecture can be replaced with distributed systems, such as heterogeneously parameterized networks of springs or resistors \cite{stern2020continual, stern2021supervised, dillavou2022demonstration}.  These systems evolve under physical dynamics to minimize some variational quantity, such as the total elastic energy or steady-state power dissipation \cite{scellier2017equilibrium}.  The steady-states of these physical systems can be interpreted as outputs which represent computations done on a set of input variables of the system, and this computation can be trained by iteratively nudging the parameters to lower the variational quantity for the desired output in response to a given input.  

\begin{figure*}[ht!]
\begin{center}
\includegraphics[width=\textwidth]{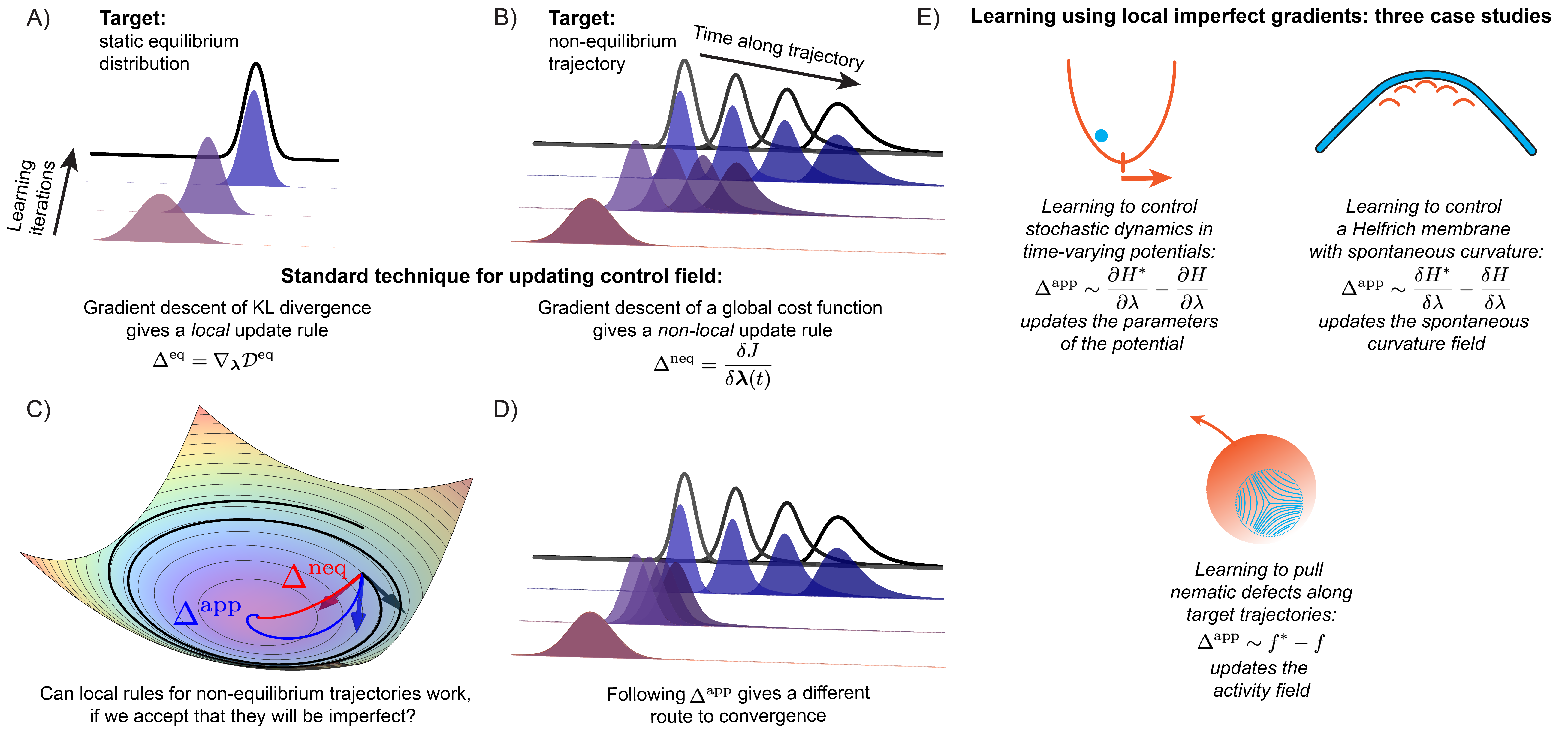}
\caption{Using imperfect gradients to solve inverse problems of non-equilibrium control.   A)  A classic learning problem is to fit the parameters of a model distribution (purple-blue) to that of a target (black).  This can be solved by flowing the parameters down the gradient of the Kullback-Leibler divergence between the two distributions.  B) 
In a dynamical setting, the objective is to fit a non-equilibrium trajectory of distributions to a given target trajectory.  This is commonly achieved using a non-local cost function $J$ of the entire trajectory.  C)  Illustration of the approximate update rule $\Delta^\text{app}$ acting as a descent direction with respect to the ``correct'' gradient $\Delta^\text{neq}$.  The black arrow has a negative dot product with $\Delta^\text{neq}$ and hence does not act as a descent direction.  D)  Same as the dynamical learning task in panel A, but illustrating that following the approximate gradient $\Delta^\text{app}$ yields a different route to convergence.  
 E) Three case studies examined in this paper.}
\label{Fig1Schematic}
\end{center}
\end{figure*}

Although recent works have studied how local update rules based on variational quantities can compute functions on static sets of data in non-equilibrium settings \cite{scellier2018extending, stern2022physical}, they have not explored how these local update rules can be used to control non-equilibrium dynamics (Figure \ref{Fig1Schematic}A).  One can draw a parallel between backpropagation for training machine learning models, which requires a global cost function and non-local propagation of information, and optimal control \cite{bechhoefer2021control}, which similarly uses numerical descent of a global cost function and requires complete model specification to work (Figure \ref{Fig1Schematic}B).  Much recent progress has been made in formulating and solving optimal control strategies for non-equilibrium dynamics, often with the aim of minimizing the heat dissipated by the trajectory \cite{shankar2022optimal, rotskoff2017geometric, iram2021controlling, chennakesavalu2023unified, gingrich2016near, das2019variational, schmiedl2007optimal, solon2018phase, davis2024active, engel2023optimal}.  Additionally, related works have successfully leveraged reinforcement learning (RL) to find control policies that guide active matter systems into desired dynamical states  \cite{chennakesavalu2021probing, falk2021learning}.   These approaches to non-equilibrium control yield highly optimized policies, but as a downside they involve biologically implausible steps such as perfect model specification, backpropagation of errors, non-local update rules based on global cost functions, or a long-term memory of previous control attempts.  We thus ask whether ideas for training machine learning systems using local, approximate update rules can be brought to bear on the problem of guiding non-equilibrium trajectories.  An answer to this question can have biological implications, helping bridge the gap between our current understanding of active matter systems and the types of regulatory feedback network that living organisms use to regulate these systems and carry out physiological functions \cite{levine2023physics}.   Additionally, it can help to generalize recent ideas for training physical materials with decentralized learning and simple, local update rules into non-equilibrium, time-varying settings \cite{stern2022physical, stern2023learning, pashine2019directed}.

Here, we introduce a set of spatiotemporally local learning rules to guide non-equilibrium systems along desired dynamical trajectories.  We make minimal use of the knowledge of the system's dynamics and parameters, instead relying on local comparisons of some (presumed accessible) observable for the system, such as its free energy density or probability density.  Our local update rules can be loosely viewed as flowing down an approximation to the gradient of an optimal control cost function and hence we dub these updates ``imperfect,'' in analogy to the above mentioned imperfect error signals that can be used in place of backpropagation to train machine learning models (Figure \ref{Fig1Schematic}C and D).  We consider both non-autonomous conservative systems (which obey detailed balance but have parameters varied at finite speed) and non-conservative systems (which break detailed balance), and we use numerical experiments to illustrate the local update rules in several examples of increasing complexity.  These include a particle trapped in a moving confining potential, a Helfrich membrane with a time-varying spontaneous curvature field, and an active nematic fluid (Figure \ref{Fig1Schematic}E).  In the last system, we demonstrate that a surprisingly simple update rule based on differences in free energy density can successfully train a spatiotemporal activity protocol to pull nematic defects along a desired trajectory \cite{shankar2022spatiotemporal}.  Taken together, the results in this paper suggest ways for using local learning rules in a broad class of non-equilibrium physical systems and machine learning models.

\begin{figure*}[ht!]
\begin{center}
\includegraphics[width= \textwidth]{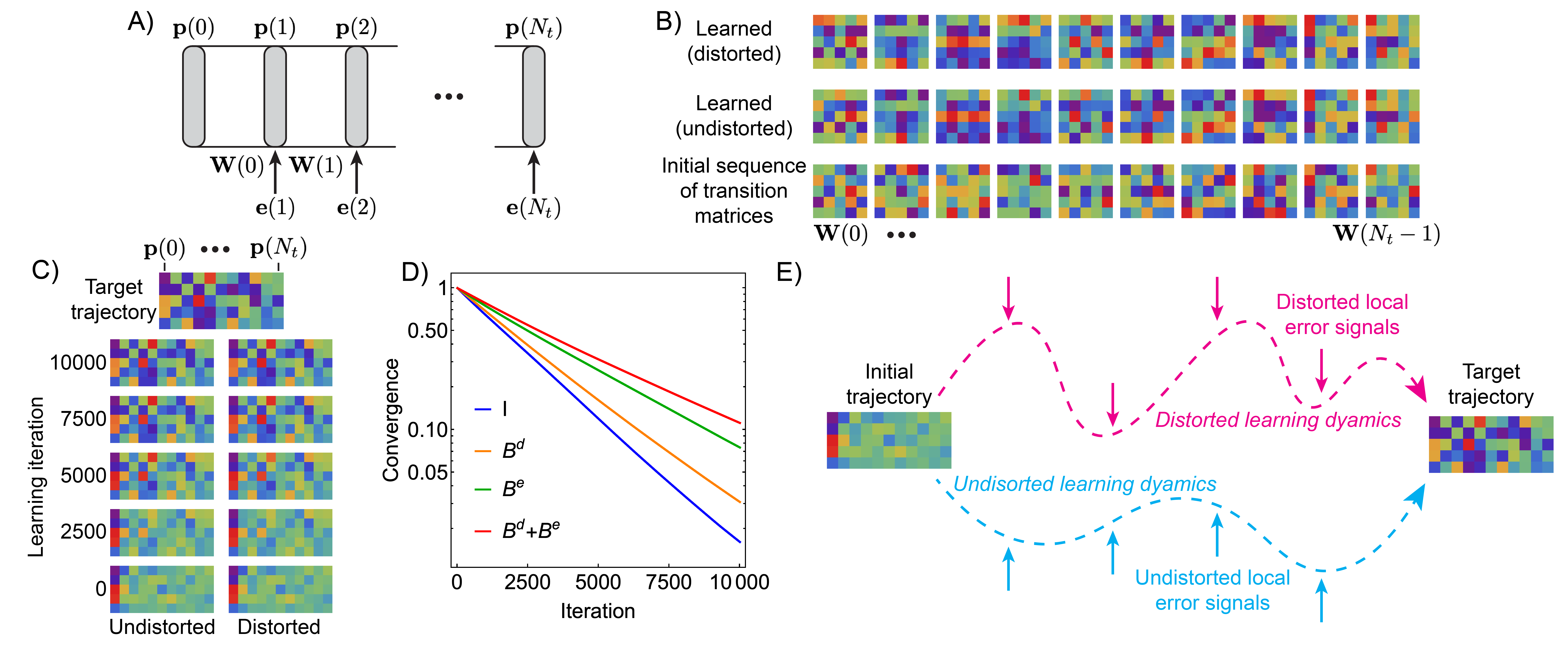}
\caption{Learning to control Markovian chain dynamics.  A) Illustration of the system in Section \ref{secMCmain} of this paper.  Gray columns indicate vectors with labels given above, and lines connecting columns indicate matrices with labels given below.  Error signals are provided according to the drawn arrows.  B)  Depiction of the sequence of transition matrices $\{\mathbf{W}(t)\}_{t=0}^{N_t-1}$, learned using distorted (with $\mathbf{B}^d$) and undistorted error signals, and staring from the same initial sequence.  Colors range from violet to red for increasing values of $W_{ij}$.  Inspection reveals that two learned sequences are not identical even though they start from the same initial sequence.  C)  Depiction of the trajectory of probability vectors $\{\mathbf{p}^n(t)\}_{t=0}^{N_t}$ obtained along the distorted and undistorted learning trajectories, for the same trial as in panel B.  Inspection reveals that the two learning processes produce final trajectories that are nearly identical to each other and match the target.  D)  Plot of the convergence, defined as $\mathcal{L}^n_T / \mathcal{L}^0_T$, as learning progresses using different error signals.  The legend indicates which distortion matrices are used for each protocol, with $I$ indicating undistorted errors.  E) Schematic illustration of the different learning trajectories which arrive at the same point of convergence.     
}
\label{TransitionMatrices}
\end{center}
\end{figure*}

\section{A primer: Learning to control non-equilibrium Markov chain dynamics with local, distorted feedback}\label{secMCmain}
Many open-loop control techniques rely on cost functions which integrate over the duration of a  dynamical trajectory and are hence global in nature \cite{bechhoefer2021control}.  We wish to explore alternative approaches in which temporally local, imperfect feedback is used instead.  As a motivating example, we first consider a Markov chain specified by a discrete sequence $\{\mathbf{W}(t)\}_{t=0}^{N_t-1}$ of stochastic transition matrices which transform an initial probability vector $\mathbf{p}(0)$ into a final vector $\mathbf{p}(N_t)$ (Figure \ref{TransitionMatrices}A).  We show that it is possible to learn a Markov chain which produces a given trajectory using temporally local error signals, and we then show that one can go further and systematically distort these local error signals without preventing convergence of the learning algorithm.  This problem is inspired by Ref.\ \citenum{lillicrap2016random}, where imperfect feedback of errors was used to avoid backpropagation when learning the parameters of a shallow neural network model (see SI Section \ref{sectransmat}A).  

We aim to learn a trajectory $\{\mathbf{p}^*(t)\}_{t=0}^{N_t}$ through probability space, assuming that $\mathbf{p}^*(0)$ is given.  We wish to avoid biologically implausible backpropagation of errors through time, so that during the $n^\text{th}$ learning iteration at time $t$ we do not use any knowledge of the errors $\mathbf{e}^n(t') = \mathbf{p}^*(t') - \mathbf{p}^n(t')$ for $t' > t+1$.  This causes our learning dynamics to be temporally local, such that each matrix $\mathbf{W}^n(t-1)$ greedily aims to connect $\mathbf{p}^n(t-1)$ to $\mathbf{p}^*(t)$ without consideration of downstream ($t'>t$) losses.  We show in SI Section \ref{sectransmat}B that even though $\mathbf{W}^n(t)$ flows down the gradient of these local loss functions $\mathcal{L}^n(t) = \frac{1}{2}(\mathbf{e}^n(t))^\intercal\mathbf{e}^n(t)$, rather than the global loss function $\mathcal{L}^n_T = \sum_{t} \mathcal{L}^n(t)$, convergence is still guaranteed over the entire trajectory.  Intuitively, this happens as the matrix $\mathbf{W}^n(0)$ first learns to connect the given $\mathbf{p}^*(0)$ to $\mathbf{p}^*(1)$, after which $\mathbf{W}^n(1)$ has the right starting point and can learn to connect $\mathbf{p}^*(1)$ to $\mathbf{p}^*(2)$, and so on.  Figures \ref{TrapResultsComposite}C and F of the next section illustrates how in practice convergence happens faster than this ``worst-case" scenario, such that later times in the trajectory have useful learning updates even before earlier times have fully converged.

We next consider the effect of systematically distorting these local loss gradients $\nabla_{\mathbf{W}(t-1)} \mathcal{L}^n(t)$ using a set of fixed random matrices $\mathbf{B}^e(t)$ and $\mathbf{B}^d(t)$.  We study two ways of doing this: one in which our error vectors are distorted as $\mathbf{e}^n(t) \rightarrow \mathbf{B}^e(t) \mathbf{e}^n(t)$, and one in which our knowledge about the dynamics is distorted as $\nabla_{\mathbf{W}(t-1)} \mathbf{p}^n(t) \rightarrow \mathbf{B}^d (\nabla_{\mathbf{W}(t-1)} \mathbf{p}^n(t))$.  In SI Sections \ref{sectransmat}C-E we show analytically and numerically that these distortions to the gradient do not prevent convergence provided that $\mathbf{B}^e(t)$ has positive eigenvalues and $\mathbf{B}^d(t)$ is positive definite.  These analytical constraints illustrate, for these Markov chain dynamics, the kinds of conditions under which local learning can work in more general dynamical settings, even in the presence of systematic errors.  We further find that these two modes of distorting the gradient have qualitatively different effects on the convergence rate:  $\mathbf{B}^e(t)$ affects convergence according to its minimum eigenvalue, which controls the slowest relaxation mode, whereas $\mathbf{B}^d(t)$ affects convergence of all modes uniformly according to its Rayleigh quotient with $\mathbf{p}^n(t-1)$.  Figures \ref{TransitionMatrices}B-E illustrate the fact that the distorted and undistorted learning dynamics both converge to the same solution, but along different paths and with a different final set of degenerate transition matrices $\{\mathbf{W}(t)\}_{t=0}^{N_t-1}$ which both effectively produce the same trajectory $\{\mathbf{p}^N(t)\}_{t=0}^{N_t}$ which matches the target $\{\mathbf{p}^*(t)\}_{t=0}^{N_t}$ (see also Figure \ref{Fig1Schematic}C).  This is similar to how the distorted feedback matrices in Ref.\ \citenum{lillicrap2016random} yield different neural network parameters which nevertheless achieve validation results comparable to those obtained from parameters trained without distortion, using full backpropagation.

The Markov chains considered here can encompass many systems of interest and motivate how a to learn dynamics with local, error-prone rules. However, Markov chains are a fairly abstract representation of physical dynamics.  We next consider how these principles of imperfect and local feedback can be used to guide dynamical trajectories using more concrete physical models.  We first consider in Section \ref{secnonauton} a class of non-equilibrium dynamics in which a conservative system is driven out of equilibrium due to non-autonomous variation of Hamiltonian parameters $\bl(\mathbf{r},t)$.  After that, in Section \ref{secnoncons} we consider systems driven by non-conservative forces that are parameterized by an activity protocol $\boldsymbol{\alpha}(\mathbf{r},t)$.  In both classes of non-equilibrium systems we take the same concrete learning problem, in which the goal is to reconstruct a target trajectory of the system $\mathbf{q}^*(\mathbf{r},t)$ by iteratively running a forward pass through the physical dynamics to generate trial trajectories $\mathbf{q}^n(\mathbf{r},t)$ and then updating the driving protocol $\bl^n(\mathbf{r},t)$ (or $\boldsymbol{\alpha}^n(\mathbf{r},t)$) using an imperfect, spatiotemporally local update rule.

\section{Learning to control conservative, non-autonomous systems}\label{secnonauton}
Learning desired dynamics in time-varying environments is a problem that is encountered in many biological and synthetic contexts \cite{levine2023physics}. For conservative systems (which obey detailed balance), the problem can be generically posed as follows. We consider a system described by a Hamiltonian $H(\mathbf{q};\bl(t))$, where $\mathbf{q}$ are the system degrees of freedom and $\bl(t)$ is a set of non-autonomously controlled parameters.  These could represent, for instance, parameters describing a fitness landscape in an immune or evolutionary context \cite{mayer2019well, schnaack2022learning, iram2021controlling}, parameters describing molecular interactions in a colloidal self-assembly context \cite{das2021variational, chennakesavalu2021probing}, or the location and stiffness of an optical trap manipulating a small particle \cite{schmiedl2007optimal,solon2018phase}.  At equilibrium with fixed $\bl$, the probability of the system obeys the Boltzmann distribution $p^\text{eq}_{\bl}(\mathbf{q}) = Z(\bl)^{-1}e^{-\beta H(\mathbf{q}; \bl)}$ where $Z(\bl)$ is the partition function and $\beta = 1/k_BT$.  Viewing $\bl$ as learnable parameters, we wish to solve the inverse problem of reconstructing a target probability \textit{trajectory}.  We assume this trajectory $p^*(\mathbf{q}, t) = p_{\bl^*(t)}(\mathbf{q})$ has been generated by evolving stochastic dynamics of the form 
\begin{equation}
    \partial_t p(\mathbf{q},t) = \mu \partial_\mathbf{q}\cdot\left(\partial_\mathbf{q}H(\mathbf{q};\bl(t)) p(\mathbf{q},t) \right) + D \partial_\mathbf{q}^2 p(\mathbf{q},t)
\end{equation}
under the protocol $\bl^*(t)$.   In these dynamics the mobility $\mu$ is related to the diffusion $D$ and inverse thermal energy $\beta$ by the Einstein relation $\mu = \beta D$.  We aim to construct a temporally local rule to iterate our guess for the protocol $\bl(t)$.

A natural cost function to consider is the Kullback-Leibler (KL) divergence $\mathcal{D}[p_{\bl^*(t)}||p_{\bl^n(t)}] \equiv \int d\mathbf{q} p_{\bl^*(t)}\ln p_{\bl^*(t)}/p_{\bl^n(t)}$ between the target trajectory $p_{\bl^*(t)}$ and the trajectory under the $n^\text{th}$ learning iteration $p_{\bl^n(t)}$.  The difficulty in computing the gradient of this cost function with respect to the parameters $\bl$ is due to the fact that when the system is not at equilibrium with the parameters $\bl(t)$, then $p_{\bl(t)}$ deviates from the Boltzmann distribution and the gradient may not have a tractable form.  Formally writing the non-equilibrium distribution in a Boltzmann-like form $p_{\bl^n(t)} \propto e^{-\beta \tilde{H}(\mathbf{q}; \bl^n(t))}$, the local gradient of the KL divergence can be expressed as
  \begin{eqnarray}
    \Delta^\text{neq} &\equiv&  \frac{ \partial \mathcal{D}[p_{\bl^*(t)}||p_{\bl^{n}(t)}]}{\partial \bl } \nonumber \\
    &=& \llangle \beta \frac{\partial \tilde{H}(\mathbf{q}; \bl^{n}(t))}{\partial \bl } \rrangle_{p_{\bl^*(t)}} - \llangle \beta \frac{\tilde{H}(\mathbf{q}; \bl^{n}(t))}{\partial \bl} \rrangle_{p_{\bl^{n}(t)} } \label{eqUpNeq}
\end{eqnarray}
where $\tilde{H}(\mathbf{q}; \bl^n(t))$ is the generally unknown exponential weight of $p_{\bl^n(t)}$.  We use the shorthand notation $\partial_\bl H(\mathbf{q}; \bl^{n}(t)) \equiv \partial_\bl H(\mathbf{q}; \bl)|_{\bl = \bl^n(t)}$.  As elaborated in SI Section \ref{secSInonatuon}A, if the protocols are quasi-static then this difficulty of determining $\tilde{H}$ disappears, and the gradient simplifies to
\begin{eqnarray}
   \Delta^\text{eq} &\equiv&  \frac{ \partial \mathcal{D}[p^\text{eq}_{\bl^*(t)}||p^\text{eq}_{\bl^{n}(t)}]}{\partial \bl } \nonumber \\
    &=& \llangle \beta \frac{\partial H(\mathbf{q}; \bl^{n}(t))}{\partial \bl } \rrangle_{p^\text{eq}_{\bl^*(t)}} - \llangle \beta \frac{H(\mathbf{q}; \bl^{n}(t))}{\partial \bl} \rrangle_{p^\text{eq}_{\bl^{n}(t)} } \label{eqUpEq}
\end{eqnarray}
where $\tilde{H}$ has been replaced with the known Hamiltonian $H$. 
The quasi-static assumption thus allows breaking the dynamical problem into a set of independent problems to which one can apply standard contrastive learning techniques based on equilibrated distributions, allowing use of $H$ in the gradient.  In SI Section \ref{secSInonatuon}B, we illustrate this further using the KL divergence evaluated over path probabilities as an alternative cost function.  

\begin{figure*}[ht!]
\begin{center}
\includegraphics[width=\textwidth]{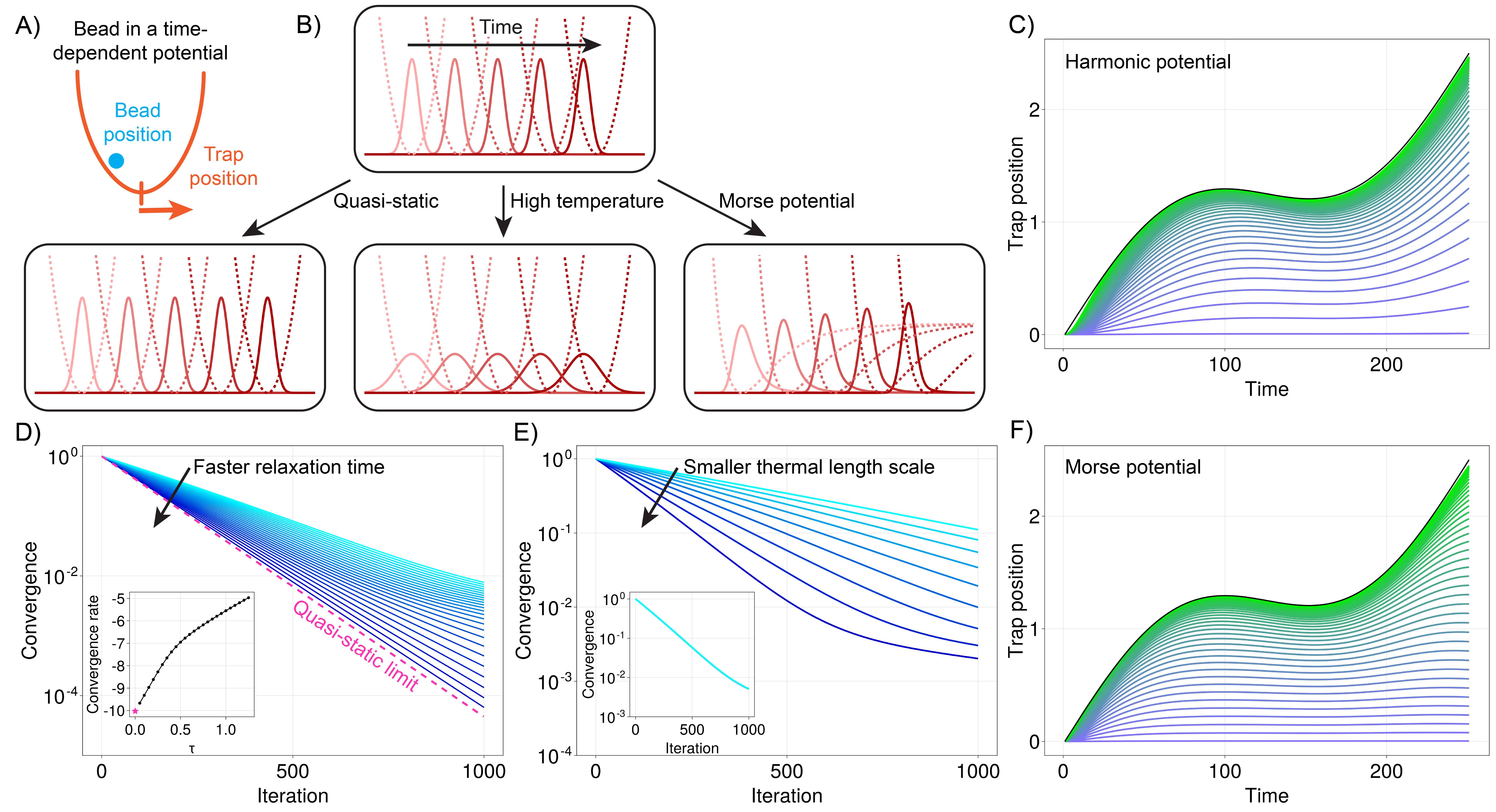}
\caption{Imperfect learning for a bead in a time-dependent potential.  A) Schematic illustration of the bead-trap system.  B)  Schematic plots of sequential probability distributions (solid lines) describing the bead location as the trap potentials (dashed lines) are moved to the right at finite speed, creating a lag between the bead and the trap.  Variations of this process are illustrated as described in the main text.  C)  The learned trap position as function of time and learning iteration, with learning rate $\eta = 0.005$ and thermal relaxation time $\tau = 0.2$.  Color ranges from purple to green as every $25^\text{th}$ learning iteration is shown.  The black line indicates the target trajectory.  D)  Convergence for the $n^\text{th}$ iteration, defined as the relative KL divergence $\int dt \mathcal{D}[p^*(t)||p^n(t)] / \int dt \mathcal{D}[p^*(t)||p^0(t)]$ is shown as $\tau$ is varied from $0.05$ (blue) to $1.25$ (cyan) in steps of $0.05$.  The pink dashed line displays the quasi-static limit obtained from learning the parameters of a single equilibrium distribution, rather than of a non-equilibrium trajectory (cf. Figure \ref{Fig1Schematic}A).  The inset shows the exponential convergence rate for each value of $\tau$ and the quasi-static limit.  E)  Convergence as the learning rate is scaled as $\eta / l^2$ for thermal length scales $\l$ ranging from $0.8$ (blue) to $1.6$ (cyan) in steps of $0.1$.  The inset displays the same, for the case when the learning rate is not scaled by $l^2$.  The nine curves corresponding to different values of $l$ are all overlapping.  F)  Same as panel C but using the Morse potential and with $\eta = 0.01$.}
\label{TrapResultsComposite}
\end{center}
\end{figure*}

If the parameters are changed at finite speed (i.e., not quasi-statically) then a \textit{lag} develops between the non-equilibrium distribution $p_{\bl(t)}(\mathbf{q})$ and the instantaneous equilibrium distribution $p^\text{eq}_{\bl(t)}(\mathbf{q})$ \cite{vaikuntanathan2009dissipation}.  Despite this lag, as an approximation to Equation \ref{eqUpEq} we can replace averages over the quasi-static distributions with averages over the actual non-equilibrium distributions:
\begin{equation}
    \Delta^\text{app} \equiv \llangle \beta \frac{\partial H(\mathbf{q}; \bl^n(t))}{\partial\bl} \rrangle_{p_{\bl^*(t)}} - \llangle \beta \frac{\partial H (\mathbf{q}; \bl^n(t))}{\partial \bl} \rrangle_{p_{\bl^{n}(t)} }. \label{eqdeltaappmain}
\end{equation}
This update can also be obtained from Equation \ref{eqUpNeq} by replacing the unknown weight $\tilde{H}(\mathbf{q};\bl)$ with the Hamiltonian $H(\mathbf{q};\bl)$.  One can view Equation \ref{eqdeltaappmain} as representing the difference between the thermodynamic force $-\partial H/\partial \bl$ expected from the target trajectory and that experienced during the $n^\text{th}$ trial.  This difference is used to inform the update to $\bl^n(t)$, causing the work increment $(\partial H/\partial \bl) \cdot \dot{\bl}^n(t)dt$ done during the $n^\text{th}$ trial to approach that done at the same time during the target trajectory.  Although $\Delta^\text{app}$ is imperfect, it has the correct fixed point at $p_{\bl^*(t)} = p_{\bl^n(t)}$.  If $\Delta^\text{app}$ additionally has a positive overlap with $\Delta^\text{neq}$, then it will act as a descent direction \cite{nocedal1999numerical} with respect to $\Delta^\text{neq}$ and will converge to the shared fixed point (Figure \ref{Fig1Schematic}C).  In SI Section \ref{secSInonatuon}C we consider a system that is close to equilibrium, so that it can be treated in the framework of linear response theory, to show that flows down $\Delta^\text{app}$ indeed converge to $\bl^*(t) = \bl^n(t)$ within a neighborhood of the fixed point.  

We emphasize that the gradient $\Delta^\text{app}$ is temporally local, in that no information of the future ($t'>t$) effects of changing $\bl(t)$ are needed to use the update rule
\begin{equation}
    \bl^{n+1}(t) \leftarrow \bl^{n}(t) - \eta \Delta^\text{app}, \label{equpdate}
\end{equation}
where $\eta$ is a scalar learning rate.  We next demonstrate the feasibility of using this update rule for a simple physical system.

\subsection{Case study: Bead in a time-dependent potential}
Here we use Equation \ref{equpdate} to learn a protocol for pulling a bead with a movable harmonic trap at position $\lambda(t)$ with stiffness $k$ (Figure \ref{TrapResultsComposite}A).  We non-dimensionalize the Fokker-Planck dynamics using the thermal length scale $l = (\beta k)^{-1/2}$ and the relaxation time $\tau = 1 / k\mu$, redefining $p  l \rightarrow p$, $q / l \rightarrow q$, $\lambda / l \rightarrow \lambda $, and $t/\tau \rightarrow t$.  This 1D system is a simple illustrative example of a broader class of systems whose degrees of freedom $\mathbf{q}$ are subject to linear forces $\mathbf{K} (\mathbf{q} - \mathbf{a})$.  Analytical expressions for these linear systems are possible because if the system starts in equilibrium then, due to the linearity of the driving force, the non-equilibrium distribution remains Gaussian for all time, having a mean and covariance matrix which lag behind their quasi-static counterparts.   In SI Section \ref{secSInonatuon}D we consider this class of systems in detail and evaluate the various updates $\Delta^\text{eq}$, $\Delta^\text{neq}$, and $\Delta^\text{app}$ explicitly in terms of these lagged quantities. 

We generate a target trajectory in the 1D bead system by moving the trap position $\lambda^*(t)$ as a function of time, and we study several variations of this process (Figure \ref{TrapResultsComposite}B).  An example trap trajectory is shown as the black line in Figure \ref{TrapResultsComposite}C, and the purple and green lines represent iterations of the learning process in which the trial trap position $\lambda^n(t)$ is updated locally in time according to Equation \ref{equpdate}.  

The relaxation time $\tau$ controls the degree to which the non-equilibrium distribution $p_{\lambda(t)}(q)$ lags behind the quasi-static distribution $p^\text{eq}_{\lambda(t)}(q)$.  Because we are not changing the stiffness parameter $k$, the updates $\Delta^\text{neq}$ and $\Delta^\text{app}$ are in fact equal for this linear system (see SI Section \ref{secSInonatuon}D).  However, convergence slows as $\tau$ increases even using the correct update $\Delta^\text{neq}$, because the typical update step for $\lambda^n(t)$ is smaller when the system is lagging (Figure \ref{TrapResultsComposite}D).  Hence for a fixed $\eta$, the more the dynamics are out of equilibrium, the slower the convergence.  

The thermal length scale $l$ controls the breadth of the probability distributions, and we now argue the effect of this parameter on the learning dynamics is something what we can trivially compensate for and thus neglect.  Dimensional analysis suggests that the learning rate $\eta$ should be divided by $\l^{2}$ in the non-dimensionalized scheme for updating the trap position $\lambda(t)$.  In Figure \ref{TrapResultsComposite}E we illustrate that as $l$ decreases and $\eta$ correspondingly increases, then the measured convergence rate indeed grows.  This growth of the non-dimensional $\eta$ with decreasing $l$ can be understood from the fact that the KL divergence cost function grows to infinity when comparing two distributions which pass into Dirac delta distributions (as $\beta\rightarrow \infty$) centered on different means.  Thus, for lower temperatures our non-dimensional learning step sizes down the gradient of this cost function will be larger because the cost function itself is larger.  If, however, we we do not scale $\eta$ by $l^{2}$ and we normalize the KL divergence by its value in the first trial to measure convergence, then the dependence of convergence rate on $\l$ drops out entirely (inset of Figure \ref{TrapResultsComposite}E). As a result, it is reasonable to neglect the dependence of the learning dynamics on temperature and consider zero-temperature dynamics in which we do not scale $\eta$ by $l^2$.  

Finally, we ask whether the approximate update rule Equation \ref{equpdate} works for a more complex non-linear example, when $\Delta^\text{neq} \neq \Delta^\text{app}$.  In place of a pure harmonic potential we use a Morse potential $H(q;\lambda, k) = (1-e^{-\sqrt{k/2}(q - \lambda)})^2 + \frac{1}{2}k_w(q - \lambda)^2$.  We add to this Hamiltonian a harmonic potential with a weak spring constant $k_w$ so that the probability density remains compactly supported.  In Figure \ref{TrapResultsComposite}F we show that the learning process in this case also converges, albeit at a slower rate than for the harmonic trap.

\subsection{Case study: Helfrich membrane with spontaneous curvature}

We next consider a system whose configuration is described by a spatially extended \textit{field} $q(r,t)$, i.e., a function of time $t$ and of a 1D spatial coordinate $r \in [0,L]$.  As a concrete example, we study a membrane system whose height $q(r,t)$ evolves to relax a Helfrich Hamiltonian (Figure \ref{HelfrichResultsComposite}A). 
 In the Monge gauge and for small $q$, the leading order contribution to the energy is $H[q(r)] = \int_0^L dr h(q(r);\lambda(r))$ where \cite{gozdz2001shape, agrawal2009calculation}
\begin{equation}
    h(q(r);\lambda(r)) =\left(\frac{\sigma}{2} +  \frac{\kappa \lambda(r)^2}{4} \right) \left(\partial_r q \right)^2 + \frac{\kappa}{2}\left(\partial_r^2 q - \lambda(r) \right)^2.
\end{equation}
Here, $\sigma$ is a surface tension, $\kappa$ is a bending rigidity, and $\lambda(r,t)$ is a spontaneous curvature field which sets the local rest value of $\partial_r^2 q$.  Because of the previously discussed trivial dependence on temperature (Figure \ref{TrapResultsComposite}E), we neglect noise here in the overdamped dynamics $\partial_t q(r,t) = -\mu \delta H /\delta q$.

\begin{figure*}[ht!]
\begin{center}
\includegraphics[width=\textwidth]{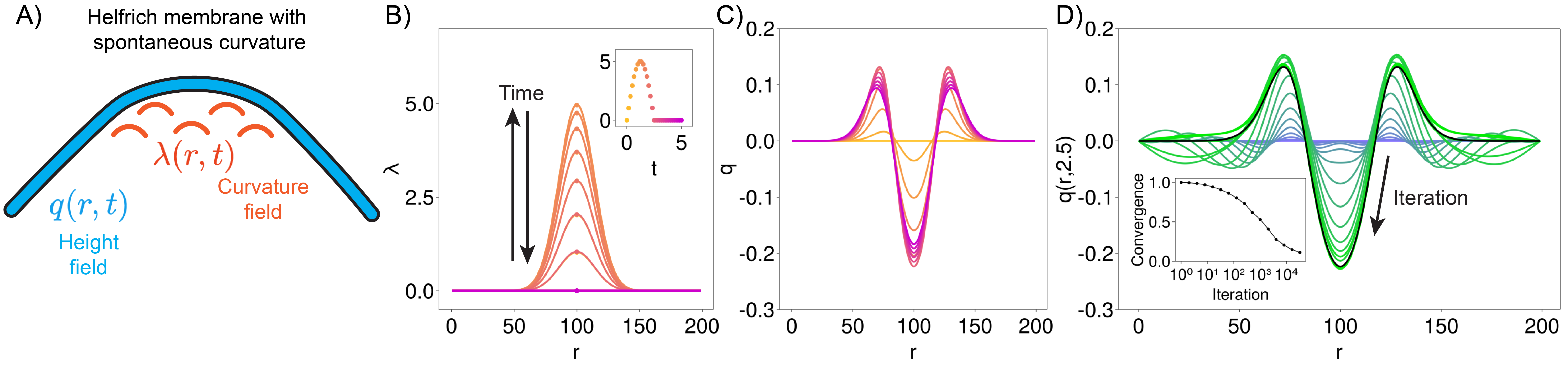}
\caption{Imperfect learning for a Helfrich membrane driven by a field of spontaneous curvature.  A)  Schematic illustration of the Helfrich membrane system.  B)  Plots of the target spontaneous curvature protocol $\lambda^*(r,t)$ at various times $t$.  The inset displays $\lambda(100,t)$ as a function of $t$, corresponding to the dots in the main plot.  C)  Plots of the height field trajectory $q^*(r,t)$ resulting from $\lambda^*(r,t)$ in panel B.  D)  Plots of the height field $q^*(r,2.5)$ as the learning iterations increase (from purple to green).  The inset shows the convergence (defined as $\int dr dt|q^*(r,t) - q^n(r,t)| / \int dr dt|q^*(r,t) - q^0(r,t)|$)  as a function of iteration. We used $\eta = 0.25$, $\beta = 1$, $\mu = 1$, and $\kappa = \sigma = 10$ (in simulation units) for these results.}
\label{HelfrichResultsComposite}
\end{center}
\end{figure*}

We take the spontaneous curvature field $\lambda(r,t)$ to be externally controllable, for example due to a spatiotemporal protocol of curvature-inducing proteins that bind to a lipid membrane \cite{zimmerberg2006proteins}.  Varying $\lambda(r,t)$ at finite speed drives the membrane out of its initially equilibrated flat configuration $q(r,0) = 0$ and generates a non-equilibrium height field trajectory $q(r,t)$.  

As in the previous example, we specify a target protocol $\lambda^*(r,t)$ and generate its corresponding height field trajectory $q^*(r,t)$ (Figure \ref{HelfrichResultsComposite}B and C).  We then iteratively learn the target protocol starting from an initial guess, $\lambda^0(r,t) = 0$, by incrementing $\lambda^n(r,t)$ using spatiotemporally local updates which are approximations to an inaccessible non-equilibrium gradient. Due to the continuous space dimension of this problem, we use the functional gradients
$\delta H / \delta \lambda$ in Equation \ref{equpdate}.

In Figure \ref{HelfrichResultsComposite}D we illustrate convergence of the imperfect, spatiotemporally local learning rule.  To understand how a spatially continuous domain does not prevent learning using local learning rules, we can view the amplitudes $\lambda_m(t)$ of the $m^\text{th}$ spatial Fourier mode of $\lambda(r,t)$ as separate learning degrees of freedom (see SI Section \ref{secSInonatuon}E).  In the limit of small $q(r,t)$ and $\lambda(r,t)$, the height field dynamics can be linearized so that the corresponding evolution equations for the amplitudes $q_m(t)$ decouple.  These amplitude evolution dynamics can then be mapped onto a discrete system of linear forces, for which we previously demonstrated that convergence is possible.  These considerations imply that learning membrane dynamics would be possible over a 2D spatial domain as well.

We have shown it is possible to learn non-autonomous protocols of Hamiltonian parameters driving conservative physical systems by using spatiotemporally local comparisons to a target trajectory.  While local rules work trivially in quasi-static conditions, our results demonstrate that local approximations also work in non-quasi-static conditions because they have positive overlap with a ``correct'' non-local update rule (Figure \ref{Fig1Schematic}C).  We have demonstrated that this principle holds across conditions of relaxation time and temperature, for non-linear potentials, and for spatially continuous degrees of freedom.  Having illustrated that imperfect local learning rules can work to control the dynamics of non-autonomous conservative systems, we next consider a more challenging class of systems involving non-conservative forces.   

\section{Learning to control non-conservative systems}\label{secnoncons}
A new difficulty for learning to control non-conservative systems is that even at long times the system is not guaranteed to equilibrate with respect to the current parameters of its Hamiltonian and will instead occupy a non-equilibrium steady-state.  This steady-state distribution will be characterized by an exponential weight $\Phi$ which is generally unknown, and as a result trying to minimize the KL divergence cost function will involve intractable gradients of $\Phi$ (see SI Section \ref{secSICRN}A).  However, progress can be made by considering alternative cost functions which lead to approximate gradients that involve accessible quantities.  We demonstrate this in a complex paradigmatic active matter system, active nematics, by illustrating how an imperfect learning rule can be used to guide topological defects along desired trajectories.

\begin{figure*}[ht!]
\begin{center}
\includegraphics[width=\textwidth]{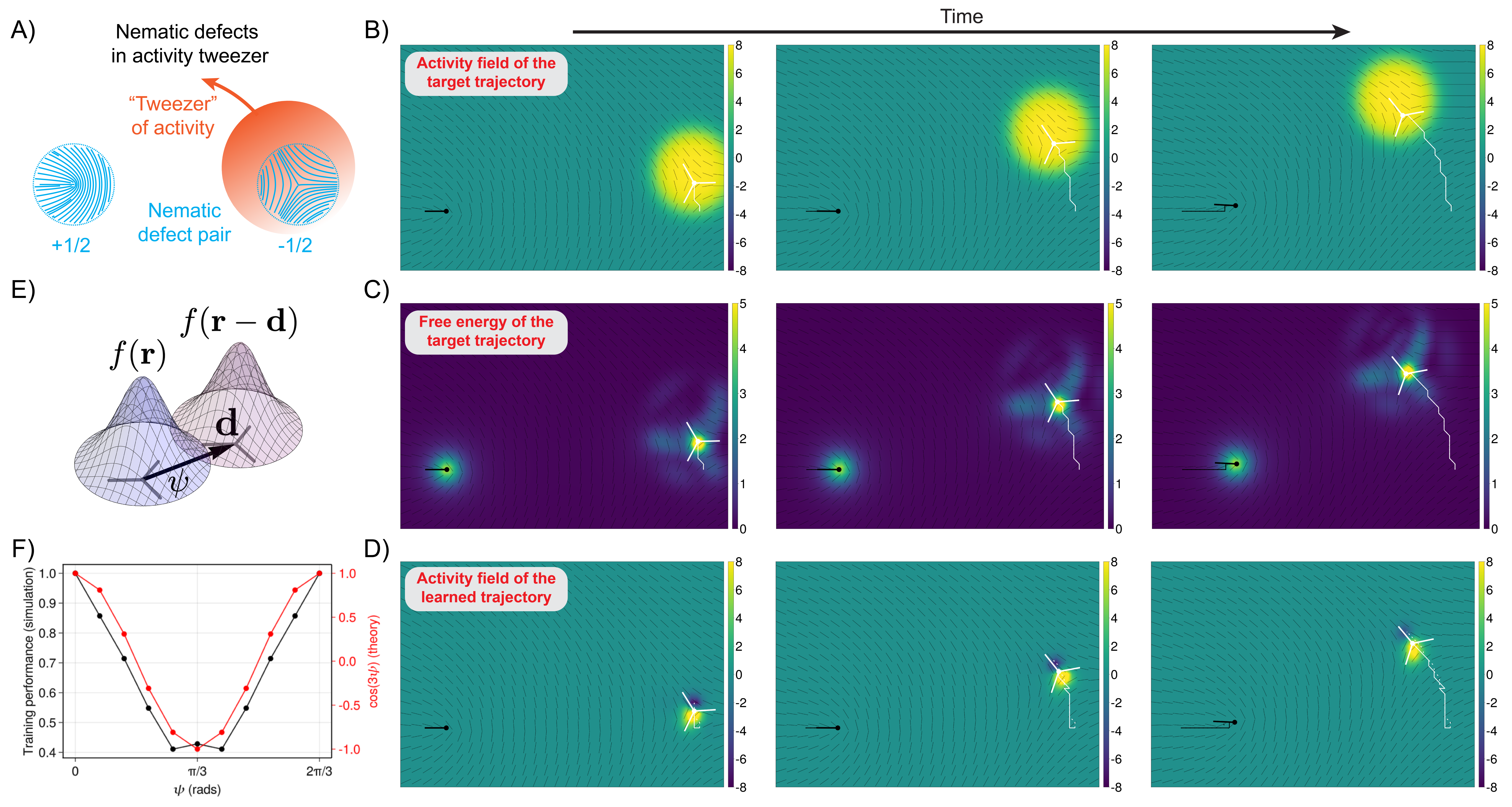}
\caption{Imperfect learning for active nematic defect dynamics.  A)  Schematic illustration of the active nematic system.  B)  Snapshots of the activity profile used to pull the $-1/2$ defect (white dot and lines) along a partial circular trajectory.  The thick black dot and lines represent the $+1/2$ defect.  The thin black lines represent the nematic field $\mathbf{Q}(\mathbf{r},t)$ and color represents $\alpha^*(\mathbf{r},t)$.  C)  Snapshots of the free energy density $f(\mathbf{r},t)$ accompanying the nematic fields in panel B.  The free energy is scaled by $10^3$.  D)  Snapshots of the trained activity protocol $\alpha^N(\mathbf{r},t)$, showing that the trained defect trajectory (solid white lines) reproduces the target trajectory (dashed white line).  The color scheme is the same as panel B.  E)  Schematic illustration of the free energy profiles under displacement of the $-1/2$ defect by a vector $\mathbf{d}$.  The gray lines indicate the defect orientation, relative to which the angle $\psi$ with respect to which $\mathbf{d}$ is defined.  F)  Plots of the simulated training performance, defined as $1 - |\mathbf{r}^*_-(T) - \mathbf{r}^N_-(T)| / |\mathbf{r}^*_-(T) - \mathbf{r}^0_-(T)|$, as the angle $\psi$ is varied.  The predicted overlap $\sim \text{cos}(3\psi)$ is plotted on the same axis.  For this comparison, short target trajectories were obtained by translating the initial nematic field $\mathbf{Q}^*(\mathbf{r},t) = \mathbf{Q}(\mathbf{r} - vt \mathbf{d}, 0)$ at a speed $v = 0.01$ to allow perfect control of the pulling direction $\mathbf{d}$.  Simulation parameters for these results are provided in SI Section \ref{secSInematic}A.} 
\label{NematicResultsComposite}
\end{center}
\end{figure*}

\subsection{Case study: Nematic defects in activity tweezer}
Finally, we consider active nematics, composed of locally extensile or contractile force dipoles that obey a liquid-crystal hydrodynamic theory \cite{doostmohammadi2018active, marchetti2013hydrodynamics}.  Active nematic dynamics describe several systems of biological interest including solutions of short biopolymers mixed with molecular motors \cite{banerjee2020actin, decamp2015orientational}, bacterial colonies \cite{copenhagen2021topological, yaman2019emergence}, and the epidermal layer of cells in developing organisms \cite{maroudas2021topological, guillamat2022integer}.   

Active nematics can be modeled using an order parameter field $\mathbf{Q}(\mathbf{r},t)$ (a symmetric and traceless tensor capturing the degree of apolar alignment of force dipoles) and a flow field $\mathbf{v}(\mathbf{r},t)$.  The evolution of $\mathbf{Q}(\mathbf{r},t)$ involves non-linear coupling to the flow field as well as a relaxational term arising from the Landau de-Gennes free energy function $F = \int d\mathbf{r}f(\mathbf{r})$ (see SI Section \ref{secSInematic}A) for details.  The flow field $\mathbf{v}$ is driven by two stress tensors, an Ericksen stress which the nematic field sets up in response to deviations from its free energy minimum, and an active stress $\boldsymbol{\sigma}^a = \alpha \mathbf{Q}$ which is a non-conservative term resulting from the activity of, for example, molecular motors walking on pairs of filaments. 

As a concrete learning problem in this system, we consider the task of manipulating the motion of \textit{defects} in the nematic field (Figure \ref{NematicResultsComposite}A).  Defects in active nematics are singular points in the $\mathbf{Q}(\mathbf{r},t)$ field around which the alignment direction rotates by a half-integer multiple of $2 \pi$ radians \cite{shankar2022topological}.  Figure \ref{NematicResultsComposite}A illustrated two types of defects, $+1/2$ and $-1/2$, which form a topologically neutral pair.  In confined or periodic systems, defects are topologically protected unless they annihilate with a defect of opposite charge.  Positioning these defects within the fluid is thought to be biologically important, as it has been demonstrated that defects in the epidermal nematic field of developing \textit{Hydra} correspond to morphological organizing centers such as the location of the future mouth \cite{maroudas2021topological, guillamat2022integer}.  Recent work illustrated that the motion of these defects can be manipulated using tightly localized gradients in the activity field $\alpha(\mathbf{r},t)$ (called ``activity tweezers'') \cite{shankar2022spatiotemporal}, similarly to how localized stresses can be created using optical tweezers to manipulate defects in colloidal crystals \cite{irvine2013dislocation}.  The necessary spatial gradients in $\alpha(\mathbf{r},t)$ can be realized $\textit{in vivo}$ through, e.g., spatial regulation of motor activating proteins \cite{banerjee2020actin}.  Additionally, it has recently become experimentally possible to exert spatial control over active nematics \textit{in vitro} by using light-activated motors or cell signaling channels \cite{linsmeier2016disordered, zhang2021spatiotemporal,lemma2023spatio,chandrasekar2023shining}.  Our method of using imperfect gradients to develop dynamical control over complex non-equilibrium systems can in principle be useful in these experimental contexts.

Using a numerical integrator of the active nematic equations of motion (whose implementation is described in Refs.\ \citenum{floyd2023simulating, floyd2022signatures, redford2024motor}), we generate target $-1/2$ defect trajectories using the activity tweezers, i.e., an activity protocol $\alpha^*(\mathbf{r},t)$, similar to those described in Ref.\ \citenum{shankar2022spatiotemporal} (Figure \ref{NematicResultsComposite}B).  Defects in the nematic field are generically characterized by persistent peaks in the free energy density $f(\mathbf{r})$ (Figure \ref{NematicResultsComposite}C).  As a result, it is reasonable to imagine that $f(\mathbf{r})$ can be used as a proxy for defect position and consider a quadratic cost function $\frac{1}{2}(f^n(\mathbf{r},t) - f^*(\mathbf{r},t))^2$.  The gradient of this cost function with respect to the learnable activity field $\alpha(\mathbf{r},t)$ involves the difficult prefactor $\partial f^n(\mathbf{r},t)/\partial \alpha(\mathbf{r},t)$.  Similarly to the manipulations in the previous sections (Equation \ref{eqalphaup}), we first assume that this prefactor can be ignored, and we consider spatiotemporally local learning rule
\begin{equation}
    \alpha^{n+1}(\mathbf{r},t) \leftarrow  \alpha^{n+1}(\mathbf{r},t) - \eta(f^*(\mathbf{r},t) -  f^n(\mathbf{r},t)). \label{eqnematicupdate}
\end{equation}
Surprisingly, this simple update rule successfully allows reconstructing the target defect trajectory (Figure \ref{NematicResultsComposite}D).  Successful implementation of this rule requires practical stabilization measures which we discuss in SI Section \ref{secSInematic}B.  We note that the learned activity protocol differs markedly from the tweezer protocol which was used to create the target trajectory.  This hints at the possibility that, while our update rule deterministically produces the same activity protocol for a given condition, there may be a degeneracy of usable activity protocols which can be accessed by alternative update rules. The update rule in Equation \ref{eqnematicupdate} produces regions of both negative (extensile) and positive (contractile) activity which is not thought to biologically realistic because an active nematic is typically either purely extensile or purely contractile, and we aim to refine this in the future.  Here we just highlight the fact that a simple free energy comparison $f^*(\mathbf{r},t) -  f^n(\mathbf{r},t)$ serves as a sufficient feedback signal to learn an activity protocol $\alpha(\mathbf{r},t)$ in a system with non-trivial physics.  

We emphasize that Equation \ref{eqnematicupdate} can be viewed as acting like an approximate gradient of a global cost function which is made difficult due to the presence of non-conservative forces, non-autonomous control parameters, and complicated non-linear hydrodynamic equations of motion.  Recent studies used top-down, global techniques such as optimal control and computation of exact coherent structures \cite{norton2020optimal, wagner2022exact} to solve similar active nematic control problems, but they require complete knowledge of the system's dynamics and parameters.  To gain intuition for why our simple local update rule works, in SI Section \ref{secSInematic}C we adapt recent analytical theory from Ref.\ \citenum{shankar2022spatiotemporal} describing the effective velocity of a $-1/2$ defect under an activity field $\alpha(\mathbf{r},t)$.  The velocity depends on second order spatial derivatives in $\alpha(\mathbf{r},t)$ via $\mathbf{v}_- = a \boldsymbol{\Sigma}:\nabla\nabla \alpha$, where $a$ is a constant prefactor and $\boldsymbol{\Sigma}$ is a certain rank-three tensor which has the appropriate symmetries to describe the orientation of the $-1/2$ defect (being invariant under a rotation by $2\pi/3$ radians).  We treat pulling the defect as a displacement of the original free energy profile $f(\mathbf{r})$ by a small vector $\mathbf{d}$, producing a new free energy profile $f(\mathbf{r} - \mathbf{d})$ (Figure \ref{NematicResultsComposite}E).  We then evaluate the Hessian of the activity $\nabla \nabla \alpha$ produced by iterating Equation \ref{eqnematicupdate} under these free energy profiles, find the velocity $\mathbf{v}_-$ which this activity field imparts to the defect, and finally evaluate the dot product between this velocity and the translation direction $\mathbf{d}$.  We find that this overlap depends as $\cos(3\psi)$ on the angle $\psi$ between the defect orientation and $\mathbf{d}$.  To test this approximate theory, we performed learning trials with various initial angles $\psi$ and measured the training performance at each angle.  Figure \ref{NematicResultsComposite}F indicates good agreement with theory.  This result explains why the approximate update rule Equation \ref{eqnematicupdate} works, and it implies that convergence depends on the orientation of the defect.  This also suggests that scalar fields besides the free energy $f(\mathbf{r})$, which might better capture details of the defect orientation, would likely serve as better sources of feedback, but we leave this to future work.

\section{Discussion}
We have demonstrated in several physical systems the idea that full top-down control is not necessary to guide complex non-equilibrium dynamics. Rather, simple, approximate (and in some cases thermodynamically motivated) update rules which make only spatiotemporally local comparisons can be used instead.  As illustrated in Figure \ref{Fig1Schematic}C, an update rule that at least somewhat aligns with a ``correct'' gradient can act as a descent direction and allow flowing down the loss landscape.  Here, we construct imperfect update rules out of easily measured observables, such as forces averaged over the trial and target distributions (Equation \ref{eqdeltaappmain}), or free energy densities (Equation \ref{eqnematicupdate}).  Ensuring convergence in each case requires some system-specific considerations, but we expect that the broader idea of using easily accessible information to guide complex dynamics in place of top-down control is general. 

The results in this paper extend the notion of imperfect gradient descent as a successful optimization strategy to the domain of dynamical control based on local error signals.  Another class of techniques in which approximated, temporally local feedback signals are used to inform updates to a control policy is temporal difference RL \cite{sutton2018reinforcement, silver2014deterministic}.  Our method bears certain conceptual similarities with this class of techniques, but with the key difference that our method does not learn a value function.  The value function in RL encodes expected rewards, and it is commonly estimated using temporally local, approximate error signals obtained through exploration of the environment.  To optimize its expected reward, the agent then updates its policy (i.e., its choice of actions) by performing gradient ascent on its estimate of the value function.  In a standard RL algorithm, neural networks for both a critic (which learns the value function from experience) and an actor (which optimizes policy based on the critic) need to be trained and stored in memory.  By contrast, our method bypasses the need to learn a value function by directly prescribing a simple, physically-motivated approximation to a value function gradient, which we call $\Delta^\text{app}$.  Thus, we leverage physical insight to provide a critic, allowing us circumvent algorithmic machinery which is typically needed in RL to provide meaningful updates to the agent's policy.  We elaborate on this comparison to RL in SI Section \ref{secSIRL}.

We expect that these methods can be of interest in several domains of research.  Relaxing the constraint of having equilibrated distributions could feasibly allow accelerating methods in generative machine learning.  For example, similar arguments were previously used to justify methods like CD-n \cite{carreira2005contrastive}.  Additionally, biological information-processing systems like the immune system operate in dynamic, non-equilibrium environments \cite{schnaack2022learning, mayer2019well}, where the principles explored in this paper might apply.  Finally, a current challenge in biology is to identify principles which living organisms might utilize to dynamically control their active mechanochemical machinery and carry out biologically useful tasks \cite{levine2023physics}.  We imagine that imperfect gradients can serve as such a principle, particularly as demonstrated in the task of active nematic defect control.  While we considered a specific inverse problem setting, we expect that one can generalize this principle to help study other biologically relevant tasks of non-equilibrium control, such as maintaining homeostasis or searching for optimal trajectories.  Additionally, recent advances in training physical materials using simple, local update rules \cite{stern2022physical, stern2023learning, pashine2019directed} can feasibly be generalized using these results to work in dynamical, non-equilibrium settings.

\section*{Acknowledgments}
We wish to thank Jordan Horowitz, Grant Rotskoff, Menachem Stern, Martin Falk, Luca Scharrer, Suraj Shankar, Jeremy Owen, Agnish Kumar Behera, and Matthew Du for helpful discussions.  This work was mainly supported by funds from DOE BES Grant DE-SC0019765 (CF and SV). ARD acknowledges support from the University of Chicago Materials Research Science and Engineering Center, which is funded by the National Science Foundation (NSF) under award number DMR-2011854 and NSF award MCB-2313725. CF acknowledges support from the University of Chicago through a Chicago Center for Theoretical Chemistry Fellowship.  The authors acknowledge the University of Chicago’s Research Computing Center for computing resources.

\end{twocolumngrid}

\clearpage

\singlespacing

\widetext
\begin{center}
	\textbf{\large Supplementary Information}
\end{center}
\setcounter{section}{0}

\section{Feedback alignment for transition matrices}\label{sectransmat}

In Ref.\ \citenum{lillicrap2016random} it is shown that a neural network can be trained without the use of exact backpropagation, in the sense that a transposed weight matrix $(\mathbf{W}^1)^\intercal$ (which would be required for exactly backpropagating errors to an upstream weight matrix $\mathbf{W}^0$) can be replaced with an random matrix $\mathbf{B}$ of full rank, and this substitution will not prevent convergence of the learning dynamics.  This surprising result is due to the fact that learning dynamics for $\mathbf{W}^1 $ cause it to eventually \textit{align} with $\mathbf{B}^\intercal$, such that $\mathbf{B}$ can pass relevant error information to $\mathbf{W}^0$.  In Ref.\ \citenum{lillicrap2016random}, this is illustrated numerically using shallow neural networks (with non-linear activation functions) and mathematically proven for shallow linear networks.  We first summarize their mathematical analysis, after which we generalize the feedback alignment idea to learn sequences of \textit{transition matrices} from trajectories.

\subsection{Summary of analysis in Ref.\ \citenum{lillicrap2016random}}
The authors of Ref.\ \citenum{lillicrap2016random} consider the linear system
\begin{eqnarray}
	y_i &=& W^1_{ij} h_j \\
	h_j &=& W^0_{jk} x_k
\end{eqnarray}
where $\mathbf{x}$ is the input data, $\mathbf{h}$ is a hidden layer, and $\mathbf{y}$ is the final output (see Figure \ref{TransitionMatrices}A of the main text).  Summation of repeated indices is implied throughout.  The correct output is labeled $\mathbf{y}^*$, and the error is defined as 
\begin{equation}
	e_i = y_i^* - y_i = \left(T_{ik} - W^1_{ij}W^0_{jk}\right)x_k \equiv E_{ik} x_k,
\end{equation} 
where $T_{ik}$ is the correct mapping from $x_k$ into $y^*_i$.  From $e_i$ the quadratic loss $\mathcal{L} = \frac{1}{2} e_k e_k$ is computed.  To minimize this loss via gradient descent, one updates $W^0_{ij}$ and $W^1_{ij}$ using the derivatives
\begin{eqnarray}
	\Delta W^0_{ij} \sim -\frac{\partial \mathcal{L}}{\partial W^0_{ij}} &=& -\frac{\partial \mathcal{L}}{\partial e_k}\frac{\partial e_k}{\partial W^0_{ij}} \nonumber \\
	&=& e_k W^1_{ki} x_j \label{eqtransmatup0}
\end{eqnarray}
and 
\begin{eqnarray}
	\Delta W^1_{ij} \sim-\frac{\partial\mathcal{L}}{\partial W^1_{ij}} &=& -\frac{\partial \mathcal{L}}{\partial e_k}\frac{\partial e_k}{\partial W^1_{ij}} \nonumber \\
	&=& e_i W^0_{jm} x_m. \label{eqtransmatup1}
\end{eqnarray}
To compute the product $e_k W^1_{ki}$, i.e. $(\mathbf{W}^1)^\intercal \cdot \mathbf{e}$, in the derivative with respect to $\mathbf{W}^0$, one needs access to the transpose of $\mathbf{W}^1$.  To avoid this transposition operation, the authors of Ref.\ \citenum{lillicrap2016random} replace $(\mathbf{W}^1)^\intercal$ by a random matrix of full rank, $\mathbf{B}$, so that
\begin{equation}
	\Delta W^0_{ij} \sim B_{ik} e_k x_j.
\end{equation}
To understand how this substitution affects the learning process, the authors of Ref.\ \citenum{lillicrap2016random} write the continuous gradient descent dynamics of $\mathbf{W}^0$ and $\mathbf{W}^1$:
\begin{equation}
	\dot{W}^0_{ij} = \eta B_{ik}E_{km}X_{mj}
\end{equation}
and 
\begin{equation}
	\dot{W}^1_{ij} = \eta E_{ik}X_{km}W^0_{jm}
\end{equation}
where $X_{ij} \equiv x_i x_j$ and the dots denote a derivative with respect to the learning iteration, and $\eta$ is a learning rate.  For an ensemble of scaled inputs $x_i$ which are drawn from $\mathcal{N}(0,1)$, the matrices $X_{ij} = \delta_{ij}$ in expectation, so that
\begin{equation}
	\dot{W}^0_{ij} = \eta B_{ik}E_{kj}
\end{equation}
and 
\begin{equation}
	\dot{W}^1_{ij} = \eta E_{ik}W^0_{jk}.
\end{equation}
Next, we imagine freezing $\mathbf{W}^1$ and consider the evolution of $\mathbf{W}^0$.  The matrix $E_{kj} = T_{kj} - W^1_{km}W^0_{mj}$ is linear in $W^0_{mj}$, and one can show that the most probable evolution of $\mathbf{W}^0$ is to grow in magnitude under these dynamics.  Hence, 
\begin{equation}
	\frac{d}{dt} W^0_{ij}W^0_{ij} = 2 \eta W^0_{ij}B_{ik}E_{kj} > 0, \label{eqtransmata1}
\end{equation}
which implies that $W^0_{ij}$ comes to align with $B_{ik}E_{kj}$.  Next, we freeze $\mathbf{W}^0$ and evolve $\mathbf{W}^1$, and consider the quantity
\begin{equation}
	\frac{d}{dt} B_{ik} W^1_{ki} = \eta B_{ik} E_{kj}W^0_{ij}.
\end{equation}
Equation \ref{eqtransmata1} implies that $\frac{d}{dt} (B_{ik} W^1_{ki}) > 0$, which means that $\mathbf{B}$ comes to align with $(\mathbf{W}^1)^\intercal$ as was to be shown.  

\subsection{Extension to chains of transition matrices}
We now build on this analysis and consider how to leverage feedback alignment to imperfectly learn a chain $\{ \mathbf{W}(t) \}_{t=0}^{N_t-1}$ of $N_t$ transition matrices, which act on probability vectors $\mathbf{p}(t)$ as
\begin{equation}
	p_{i}(t) = W_{ij}(t-1)p_j(t-1).
\end{equation}
As a probability vector over a $M$ dimensional space, $\mathbf{p}(t)$ obeys $\sum_{i=1}^M p_i(t) = 1$, and due to conservation of probability we must have 
\begin{equation}
	\sum_i W_{ij} = 1 \ \ \ \ \ \forall j, \label{eqrowsum}
\end{equation}
i.e., the columns of the transition matrices sum to unity.  Furthermore, the entries $W_{ij}$ must be non-negative in order to physically represent the transition probability from state $j$ to state $i$.  See Figure \ref{TransitionMatrices}B of the main text for an illustration of the transition matrix chain.  

We pose the problem of learning a target trajectory in probability space $\{ \mathbf{p}^{*}(t) \}_{t=0}^{N_t}$ through temporally local comparisons to trial trajectories $\{ \mathbf{p}^{n}(t) \}_{t=0}^{N_t}$, and we take the first vector $\mathbf{p}(0)$ as given and fixed.  We define the error of the $n^\text{th}$ trial at each time $t$ during the trajectory as 
\begin{equation}
	e_i^n(t) = p_i^*(t) - p_i^n(t) = E_{ik}^n(t) p_k(0)
\end{equation}
where 
\begin{equation}
	\mathbf{E}^n(t) =\prod_{t'=0}^{t'-1}\mathbf{W}^*(t') - \prod_{t'=0}^{t'-1}\mathbf{W}^n(t') \equiv \mathbf{V}^*(t-1) - \mathbf{V}^n(t-1).
\end{equation}
The matrix products here are understood to be ordered as $\mathbf{W}^*(t'-1) \mathbf{W}^*(t'-2) \ldots \mathbf{W}^*(0)$.  

We define a total loss function as 
\begin{equation}
	\mathcal{L}^n_T = \sum_{t=1}^{N_t}\mathcal{L}^n(t)  = \frac{1}{2}\sum_{t=1}^{N_t}e^n_k(t) e^n_k(t),  \label{eqLT}
\end{equation}
which sums over local loss functions $\mathcal{L}^n(t)$ defined at each time $t$.  To find transition matrices which minimize this loss, we consider gradients 
\begin{equation}
	\frac{\partial \mathcal{L}^n_T}{\partial W^n_{ij}(t-1)} = \sum_{t'=1}^{N_t}\frac{\partial \mathcal{L}^n(t')}{\partial W^n_{ij}(t-1)} = \sum_{t'=t}^{N_t}\frac{\partial \mathcal{L}^n(t')}{\partial W^n_{ij}(t-1)} \label{eqLossfulltime}
\end{equation}
where the final equality is due to causality.  For $t' > t$, we evaluate
\begin{eqnarray}
	\frac{\partial \mathcal{L}^n(t')}{\partial W^n_{ij}(t-1)} &=& -e^n_k(t') \frac{\partial p^n_k(t')}{\partial W^n_{ij}(t-1)} \nonumber \\
	&=& -e^n_k(t') \frac{\partial V^n_{km}(t'-1)}{\partial W^n_{ij}(t-1)} p_m(0) \nonumber \\
	&=& -e^n_k(t')V^n_{ki}(t'-1:t) V^n_{jm}(t-2) p_m(0) \label{eqtransdLdWij}
\end{eqnarray}
where
\begin{equation}
	\mathbf{V}^n(t'-1:t) \equiv \prod_{t''=t}^{t'-1}\mathbf{W}^n(t'').
\end{equation}
Now, any updates to $W^n_{ij}$ must preserve its column sum so it remains a stochastic matrix. A sum over $i$ in Equation \ref{eqtransdLdWij} is not guaranteed to be zero because the row sum for a product of transition matrices is arbitrary.  Additionally, the dot product $e^n_k(t')V^n_{ki}(t'-1:t)$, i.e. $(\mathbf{V}^n(t'-1:t))^\intercal \cdot \mathbf{e}^n(t')$, involves a transposition operation.  The term $V^n_{ki}(t'-1:t)$ represents backpropagation of the error signal from future times $t' > t$ to time $t$, and is hence non-local in time.  For these reasons we are motivated to neglect all derivatives $ \frac{\partial \mathcal{L}^n(t')}{ \partial W^n_{ij}(t-1)}$ for times $t' >t$ and adjust $W^n_{ij}(t-1)$ only according to the \textit{local} gradient at $t'=t$ (cf. Equation \ref{eqtransmatup1} above),
\begin{equation}
	\frac{\partial \mathcal{L}^n(t)}{\partial W^n_{ij}(t-1)} = -e^n_k(t) \delta_{ik} V^n_{jm}(t-2) p_m(0). \label{eqtransdLdWijlocal}
\end{equation}
To use this expression when $t = 1$ we set $V^n_{jm}(-1) = \delta_{jm}$.  We note that summing Equation \ref{eqtransdLdWijlocal} over $i$ yields zero, because $\sum_k e^n_k(t) = 0$, so this moving along gradient does not affect the column sum of $W^n_{ij}(t-1)$. 

We now consider how the error $\mathbf{e}^n(t)$ evolves under the learning dynamics 
\begin{equation}
	\dot{W}^n_{ij}(t-1) = - \eta \frac{\partial \mathcal{L}^n(t)}{\partial W^n_{ij}(t-1)} = \eta e_i^n(t) V^n_{jm}(t-2) p_m(0) \label{eqlearndynorig}
\end{equation}
where the dot denotes a derivative with respect to the training iteration $n$, not with respect to the trajectory time index $t$.  We have that 
\begin{eqnarray}
	\dot{e}^n_i(t) &=& \dot{E}_{ik}p_k(0) \nonumber \\
	&=& - \dot{V}^n_{ik}(t-1)p_k(0) \nonumber \\
	&=& - \left(\dot{W}^n_{im}(t-1)V_{mk}(t-2) + W^n_{im}(t-1)\dot{W}^n_{ml}(t-2) V_{lk}^n(t-3) + \ldots \right) p_k(0). \label{eqtranseqdotfull}
\end{eqnarray}
To simplify this expression, let us first consider $t=1$:
\begin{eqnarray}
	\dot{e}^n_i(1) &=& -\dot{W}^n_{ik}(0) p_k(0) \nonumber \\
	&=& -\eta e^n_i(1) p_k(0) p_k(0). \label{eqe1dyn}
\end{eqnarray}
Because the scalar quantity $a_1 \equiv p_k(0) p_k(0)$ is positive, this differential equation clearly represents an exponential decay of $e^n(1)$ to zero as $n \rightarrow \infty$ at a rate $\eta a_1$.  Next we consider $t=2$:
\begin{eqnarray}
	\dot{e}^n_i(2) &=& -\left(\dot{W}^n_{im}(1)W^n_{mk}(0) + W^n_{im}(1)\dot{W}^n_{mk}(0)\right) p_k(0) \nonumber \\
	&=& -\eta e_i^n(2)W^n_{ml}(0)p_l(0) W^n_{mk}p_k(0) + W^n_{im}(1)\dot{e}_m^n(1).
\end{eqnarray}
We have already shown that after iteration $\sim 1/\eta a_1$ the factor $\dot{e}_m^n(1)$ will be negligible and $\mathbf{W}^n(0) \rightarrow \mathbf{W}^\infty(0)$, so we can write
\begin{equation}
	\dot{e}^n_i(2) \approx -\eta e_i^n(2)W^\infty_{ml}(0)p_l(0) W^\infty_{mk}p_k(0) \equiv -\eta a_2 e_i^n(2).
\end{equation}
The quantity $a_2 = W^\infty_{ml}(0)p_l(0) W^\infty_{mk}p_k(0)$ is positive because it is of the form $\mathbf{p}^\intercal \mathbf{A}^\intercal \mathbf{A} \mathbf{p}$ and the matrix $\mathbf{A}^\intercal \mathbf{A}$ is positive definite.  Thus, the error $\dot{e}^n_i(2)$ will also decay exponentially to zero as $n \rightarrow \infty$.  Extrapolating this pattern to arbitrary $t$, we can therefore neglect all but the leading term in Equation \ref{eqtranseqdotfull} and write 
\begin{eqnarray}
	\dot{e}^n_i(t) &=& \dot{W}^n_{im}(t-1)V^\infty_{mk}(t-2) p_k(0) \nonumber \\
	&=& -\eta e_i^n(t) V_{jm}^\infty(t-2) p_m(0) V^\infty_{mk}(t-2) p_k(0) = - \eta a_t e_i^n(t).
\end{eqnarray}
This convergence process can be viewed as ``zippering,'' where early times $t$ in the trajectory converge first, after which later times converge as well.

\subsection{Learning chains of transition matrices with imperfect gradients}
In the previous subsection, it was assumed that the learning dynamics exactly follow the gradients $ \frac{\partial \mathcal{L}^n(t) }{\partial W^n_{ij}(t-1)}$.  Somewhat surprisingly, this was shown to be possible even when only considering temporally local loss functions (avoiding backpropagation in time).  We now go a step further and consider the conditions under which convergence would still be possible if we systematically \textit{distort} our local feedback signal.  Starting from Equation \ref{eqtransdLdWijlocal}, we identify at least two possible ways to introduce a distortion.  One is that we distort the error vector:
\begin{equation}
	e_k^n(t) \rightarrow B_{kl}^e(t) e_l^n.
\end{equation} 
The other is that we distort our ``dynamical knowledge'': 
\begin{equation}
	\frac{\partial V^n_{km}(t)}{\partial W^n_{ij}(t-1)} = \delta_{ik} V^n_{jm}(t-2) \rightarrow \delta_{ik} B^d_{jp}(t) V^n_{pm}(t-2).
\end{equation}
Introducing these random matrices $\mathbf{B}^e$ and $\mathbf{B}^d$, the distorted learning dynamics become (cf. Equation \ref{eqlearndynorig})
\begin{equation}
	\dot{W}^n_{ij}(t-1) = \eta B^e_{il}(t) e_l^n(t) B^d_{jk}(t)  V^n_{km}(t-2) p_m(0). \label{eqlearndyndist}
\end{equation}
We require that $\sum_{i}B^e_{il}e_l = 0$ if it is not to affect the column sum of $W^n_{ij}(t-1)$.  Considering as before the training dynamics of the first error $e^n(1)$, we have
\begin{eqnarray}
	\dot{e}^n_i(1) &=& -\dot{W}^n_{ik}(0) p_k(0) \nonumber \\
	&=& -\eta  B^e_{il}(1) e_l^n(1) B^d_{km}(t) p_m(0) p_k(0). \label{eqe1dyndist}
\end{eqnarray}
If $\mathbf{B}^d(1)$ is positive definite, then the product $a^d_1 \equiv B^d_{km}(t) p_m(0) p_k(0)$ is guaranteed to be positive, so we can write
\begin{equation}
	\dot{e}^n_i(1) = - \eta a_1^d B^e_{il}(1)e_l^n(1)
\end{equation}
with $\eta a_1 > 0$.  This differential equation will decay to zero if $\mathbf{B}^e(1)$ has positive eigenvalues (a slightly weaker condition than it being positive definite). For $t=2$ we have (assuming $\mathbf{e}(1)$ has appreciably decreased)
\begin{eqnarray}
	\dot{e}^n_i(2) \approx -\eta B^e_{il}(2)e_l^n(2)B^d_{mp}(2) W^\infty_{pl}(0) p_l(0) W^\infty_{mk}p_k(0) 
\end{eqnarray}
If $\mathbf{B}^d(2)$ is positive definite, then it can be split as $\mathbf{B}^d(2) = \left(\bar{\mathbf{B}}^d(2)\right)^\intercal\bar{\mathbf{B}}^d(2)$ for some $\bar{\mathbf{B}}^d(2)$, so that the product $a^d_2 = B^d_{mp}(2) W^\infty_{pl}(0) p_l(0) W^\infty_{mk}p_k(0)$ is of the form $\mathbf{p}^\intercal\mathbf{A}^\intercal \mathbf{A} \mathbf{p}$ and is hence positive.  Thus
\begin{equation}
	\dot{e}^n_i(2) = - \eta a_2^d B^e_{il}(2)e_l^n(1)
\end{equation}
will is guaranteed to converge if $\mathbf{B}^e(2)$ has positive eigenvalues.  Continuing this pattern, for arbitrary $t$ we have
\begin{equation}
	\dot{e}^n_i(t) = -\eta B^e_{il}(t) e_l^n(t) B^d_{mp}(2)V_{pl}^\infty(t-2) p_l(0) V^\infty_{mk}(t-2) p_k(0) = - \eta a^d_t B^e_{il}(t)e_l^n(1)e_i^n(t)
\end{equation}
which will also converge if $\mathbf{B}^d(t)$ is positive definite and $\mathbf{B}^e(t)$ has positive eigenvalues.

An alternative illustration of the convergence of these distorted gradients can be given by considering their overlap with the correct gradients.  For time $t$, we evaluate
\begin{eqnarray}
	\left(\frac{\partial \mathcal{L}^n(t)}{\partial W^n_{ij}(t-1)}\right)_\text{correct}\left(\frac{\partial \mathcal{L}^n(t)}{\partial W^n_{ij}(t-1)}\right)_\text{distorted} &=& e_i^n(t) V^n_{jm}(t-2) p_m(0) B^e_{il}(t) e_l^n(t) B^d_{jp}(2)V^n_{pl}(t-2) p_l(0) \nonumber \\
	&=& \left(e_i^n(t) B^e_{il}(t) e_l^n(t)\right) \left(V^n_{jm}(t-2) p_m(0) B^d_{jp}(2)V^n_{pl}(t-2) p_l(0)\right).
\end{eqnarray}
This expression is the product of two terms, each of the form $\mathbf{p}^\intercal\mathbf{A}^\intercal \mathbf{A} \mathbf{p}$ provided that $\mathbf{B}^e(t)$ and $\mathbf{B}^d(t)$ are both positive definite.  Under these conditions, this overlap is positive and thus the distorted gradient can serve as a descent direction of $\mathcal{L}^n(t)$ \cite{nocedal1999numerical}.  We note that this argument places the more stringent condition on $\mathbf{B}^e(t)$ that it be positive definite rather than have positive eigenvalues.  

The simultaneous conditions on $\mathbf{B}^e$ that it have positive eigenvalues and satisfy $\sum_{i}B^e_{il}e_l = 0$ are not contradictory.  For example, such a matrix can be constructed as 
\begin{equation}
	\mathbf{B}^e = \mathbf{I} + \tilde{\mathbf{C}} \label{eqBcreate1}
\end{equation}
where $\mathbf{I}$ is the identity matrix, $\mathbf{C}$ is positive definite, and the tilde operation maps a matrix as
\begin{equation}
	\tilde{C}_{ij} = C_{ij} - \frac{1}{M}\sum_{k=1}^MC_{kj} \label{eqBcreate2}
\end{equation}
so that its column sums are zero.  

\subsection{Effect of imperfect gradients on convergence}
The matrices $\mathbf{B}^e(t)$ and $\mathbf{B}^d(t)$ affect the rate of convergence of the error in two qualitatively different ways.  For simplicity, we focus on the effect on the first error $\mathbf{e}^n(1)$ due to $\mathbf{B}^e(1)$ and $\mathbf{B}^d(1)$.  The undistorted dynamics of $\mathbf{e}^n(1)$ are given in Equation \ref{eqe1dyn}, which represents exponential convergence of each component $e_i^n(1)$ to zero at rate $\eta a_1$.  The distorted dynamics are given in Equation \ref{eqe1dyndist}, which introduce two new features.  

First, the scalar quantity $a_1^d$ differs from $a_1$ by the Rayleigh quotient
\begin{equation}
	\frac{a_1^d}{a_1} = \frac{\mathbf{p}^\intercal(0) \mathbf{B}^d(1)\mathbf{p}(0) }{\mathbf{p}^\intercal(0)\mathbf{p}(0)} = \frac{\sum_{i=1}^M (y_i^d(1))^2 \lambda_i^d(1) }{\sum_{i=1}^M (y_i^d(1))^2}
\end{equation}
where $\lambda_i^d$ are the eigenvalues of $\mathbf{B}^d(1)$ and $y_i^d(1)$ are the components of $\mathbf{p}(0)$ in the eigenbasis of $\mathbf{B}^d(1)$.  Clearly the Rayleigh quotient quantity will depend on the the projection of initial point $\mathbf{p}(0)$ onto the eigenmodes of $\mathbf{B}^d(1)$.  Assuming random data, we can take these projections to be equal on average, so that $y_i^d(1) = y^d(1)$ for each $i$.  The quotient then simplifies to 
\begin{equation}
	\frac{a_1^d}{a_1} = \frac{1}{M} \sum_{i=1}^M \lambda_i^d(1) 
\end{equation}
which is the average eigenvalue of $\mathbf{B}^d(1)$.  This implies that including a distortion matrix $\mathbf{B}^d(1)$ can in fact accelerate convergence by effectively causing the learning dynamics to take bigger step sizes per iteration, even if the step is not directed exactly down the gradient.  If we normalize $\mathbf{B}^d(1) \rightarrow \mathbf{B}^d(1) / \lambda_\text{max}^d$(1) so that its maximum eigenvalue $\lambda_\text{max}^d$ is one, then the effect of including $\mathbf{B}^d(1)$ will be to strictly slow down convergence because $a^d_1/a_1 \leq 1$.  We emphasize that $a^d_1/a_1$ affects the convergence rate of each component of $e_i^n(1)$ by the same amount.  

By contrast, the effect second effect on the converge rate, caused by $\mathbf{B}^e(1)$, will be different for each component of $e_i^n(1)$.  Setting $a_1^d=a_1$ here, we compare
\begin{equation}
	\dot{e}^n_i(1) = -\eta a_1 e_i^n(1) \label{eqe1simp}
\end{equation}
and 
\begin{equation}
	\dot{e}^n_i(1) = -\eta a_1 B_{ik} e_k^n(1). \label{eqe1mat}
\end{equation} 
Under Equation \ref{eqe1simp} each component of $\mathbf{e}^n(1)$ will decay at the same rate, $\eta a_1$.  Under Equation \ref{eqe1mat}, the component of $\mathbf{e}^n(1)$ along the $i^\text{th}$ eigenvector of $\mathbf{B}^e(1)$ will decay at a rate $\eta a_1 \lambda_i^e$, where $\lambda_i^e$ associated eigenvalue.  At long training times, decay of $\mathbf{e}^n(1)$ will be dominated by the smallest eigenvalue $\lambda_\text{min}^e$.  
To summarize, the effect of $\mathbf{B}^d(1)$ is to alter the convergence rate of each component $e_i^n(1)$ by the same factor, equal (under the assumption of random data) to the \textit{average} eigenvalue of $\mathbf{B}^d(1)$.  On the other hand, the effect of $\mathbf{B}^e(1)$ is to alter the convergence rate of each component $e_i^n(1)$ differently according to the projection of $\mathbf{e}^n(1)$ on to the eigenbasis of $\mathbf{B}^e(1)$, and this effect will eventually be dominated by the \textit{smallest} eigenvalue of $\mathbf{B}^e(1)$.  This qualitative argument extends straightforwardly to arbitrary times $t$.

\subsection{Numerical results}
Here we numerically test the training protocols described above.  A remaining issue to address first is that, while the gradients we have introduced will ensure that the column sums of the transition matrices do not differ from one during training, they do not prevent their values from going negative.  At least two possibilities exist for handling this.  One is to include inequality constraints as Lagrange multipliers in an augmented cost function \cite{nocedal1999numerical}.  The other is to take small steps along the gradient of the unconstrained cost function and then project back onto the manifold of allowed transition matrices.  The space of matrices whose entries are in the interval $[0,1]$ and whose columns sum to one is a simplex, and algorithms exist to project data onto simplices \cite{chen2011projection}.  For simplicity, we use here a crude two-step projection 
\begin{eqnarray}
	W_{ij} & \rightarrow & W^\text{clip}_{ij} \equiv \text{max}\left(W_{ij},0\right) \ \ \ \ \ \ \text{(clip to positive values)} \\
	W^\text{clip}_{ij} & \rightarrow & \frac{W^\text{clip}_{ij} }{\sum_{k=1}^MW^\text{clip}_{kj}} \ \ \ \ \ \ \ \ \ \ \ \ \ \ \ \ \ \ \text{(renormalize the column sum to one)}. \label{eqrenorm}
\end{eqnarray}
We find in practice that this suffices to ensure that $W_{ij}$ remains a legal transition matrix.

We generate target trajectories by creating $N_t+1$ vectors $\mathbf{p}^*(t)$.  We create $\mathbf{p}^*(t)$ by drawing a random number uniformly from $[0,1]$ for each component $p_i^*(t)$, and then dividing each component by $\sum_{i=1}^Mp_i^*(t)$ so that $\mathbf{p}^*(t)$ is a probability vector.  We similarly create initial guesses for the $N_t$ transition matrices $\mathbf{W}^0(t)$ by drawing each element uniformly from $[0,1]$ and then applying Equation \ref{eqrenorm} to make the column sums equal to one.

To create the $N_t$ matrices $\mathbf{B}^d(t)$, we first create $\bar{\mathbf{B}}^d(t)$ by drawing elements uniformly from $[0,1]$, and we then set $\mathbf{B}^d(t) = \bar{\mathbf{B}}^d(t)^\intercal \bar{\mathbf{B}}^d(t)$ to ensure that it is positive definite.  We finally normalize $\mathbf{B}^d(t) \leftarrow \mathbf{B}^d(t) / \lambda_\text{max}^d(t)$ so that its maximum eigenvalue is equal to one.  To create the $N_t$ matrices $\mathbf{B}^e(t)$, we first create $\bar{\mathbf{C}}(t)$ by drawing elements uniformly from $[0,1]$, and we then set $\mathbf{C}(t) = \bar{\mathbf{C}}(t)^\intercal \bar{\mathbf{C}}(t)$.  We then apply Equations \ref{eqBcreate1} and \ref{eqBcreate2} to form $\mathbf{B}^e(t)$.  Finally we set $\mathbf{B}^e(t) \leftarrow \mathbf{B}^e(t) / \lambda_\text{max}^d(t)$ so that its maximum eigenvalue is equal to one.  

\begin{figure}[h!]
	\begin{center}
		\includegraphics[width=1.0 \textwidth]{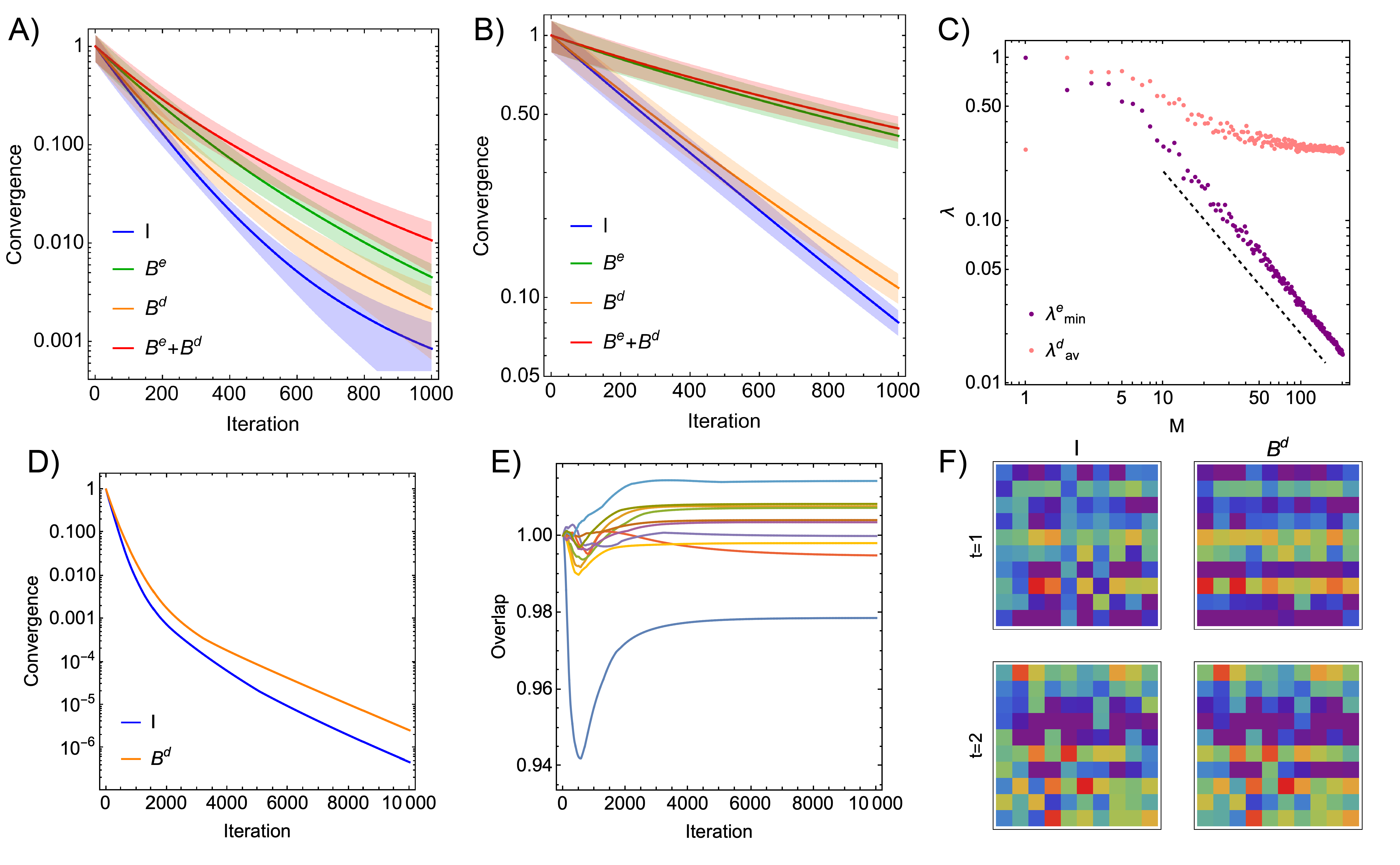}
		\caption{Numerical results for training transition matrices.  A) Convergence of the four protocols with $M = 5$.  Convergence is defined as $\mathcal{L}^n_T / \mathcal{L}^0_T$ (cf. Equation \ref{eqLT}).  Solid lines represent the mean over 10 trials and the filled curves represent the standard deviation.  B)  Convergence of the four protocols with $M = 20$.  C) Samples of the minimum eigenvalue $\lambda^e_\text{min}$ of $\mathbf{B}^e$ and average eigenvalue $\lambda^e_\text{av}$ of $\mathbf{B}^d$ as a function of $M$.  The dashed line indicates a scaling of $M^{-1}$.  D)  Convergence plots of a single trial with $M=10$ for $10,000$ training steps.  E) Overlaps for $t = 1\ldots10$ shown as different colors, with overlap defined as $W^{n,\text{I}}_{ij}(t) W^{n,d}_{ij}(t) / W^{n,\text{I}}_{ij}(t) W^{n,\text{I}}_{ij}(t)$ for transition matrices under protocols using no distortion (I), and $\mathbf{B}^d$ ($d$).  F)  Plots of the final transition matrices $\mathbf{W}^{N, \text{I}}(1)$, $\mathbf{W}^{N, d}(1)$, $\mathbf{W}^{N, \text{I}}(2)$, $\mathbf{W}^{N, d}(2)$.  Colors range from violet to red for increasing values of $W_{ij}$. }
		\label{TransitionMatrixComposite}
	\end{center}
\end{figure}

We first set $N_t = 10$, $\eta = 0.025$, and $M = 5$ and run 10 trials of training using different random initial conditions for each trial.  We try four protocols: using just $\mathbf{B}^e(t)$, using just $\mathbf{B}^d(t)$, using both, and using neither.  For each protocol we use the same initial guesses $\mathbf{W}^0(t)$.  The results are shown in SI Figure \ref{TransitionMatrixComposite}A.  The fastest convergence is achieved when no distortion is introduced, but all four protocols show convergence toward zero error.  Increasing the space dimensionality to $M = 20$, we find that distortion using $\mathbf{B}^d(t)$ performs much better than using $\mathbf{B}^e(t)$ (SI Figure \ref{TransitionMatrixComposite}B).  This can be explained due to our previous argument that convergence with $\mathbf{B}^e(t)$ depends on its slowest eigenvalue, whereas convergence with $\mathbf{B}^d(t)$ depends on its average eigenvalue.  For random matrices of increasing dimensionality $M$, the minimal eigenvalue falls more sharply with $M$ than its average eigenvalue does, as can be seen numerically (SI Figure \ref{TransitionMatrixComposite}C).  

We find that for later iterations the convergence deviates from pure exponential decay (Figure \ref{TransitionMatrixComposite}D), which can be due both to set of different relaxation timescales $\eta a_t$ for each $t$ and due to the operation of projecting back onto the simplex of legal transition matrices.  We further find that, even as both protocols using $\mathbf{B}^d$ and no distortion reach low values of the loss function, they do not learn exactly the same transition matrices (SI Figures \ref{TransitionMatrixComposite}E and F).  There is a manifold of degenerate transition matrices, each of which map $\mathbf{p}^*(t)$ into $\mathbf{p}^*(t+1)$, and the learning dynamics induced by distortion reach a different final matrix than in the un-distorted dynamics.

\newpage
\section{Imperfect learning of conservative, non-autonomous dynamics}\label{secSInonatuon}

\subsection{Lagged KL divergences}
A standard approach for training generative machine learning models is to minimize the KL divergence between a parameterized trial distribution and a target distribution.  Denoting the target distribution $p^*(\mathbf{q})$ and the trial distribution $p_\bl(\mathbf{q})$, which depends on some parameters $\bl$, one aims to minimize 
\begin{equation}
	\mathcal{D}[p^*||p_\bl] \equiv \int d\mathbf{q} p^*(\mathbf{q})\ln\frac{p^*(\mathbf{q})}{p_\bl(\mathbf{q})}.
\end{equation}
If $p_{\boldsymbol{\lambda}}(\mathbf{q})$ is an equilibrium canonical distribution with respect to a Hamiltonian $H(\mathbf{q}; \boldsymbol{\lambda})$ with $\beta = 1/k_B T$, we can write
\begin{equation}
	p_\bl(\mathbf{q}) = Z(\boldsymbol{\lambda})^{-1} e^{-\beta H(\mathbf{q}; \bl)}
\end{equation}
where $Z(\boldsymbol{\lambda})$ is the partition function.  After a few lines of algebra one can express the gradient of the KL divergence with respect to $\bl$ as
\begin{equation}
	\frac{\partial \mathcal{D}[p^*||p_{\boldsymbol{\lambda}}]}{\partial \bl } = \llangle\beta \frac{\partial H(\mathbf{q}; \bl)}{\partial \bl } \rrangle_{p^*} - \llangle\beta \frac{\partial H (\mathbf{q}; \bl)}{\partial \bl} \rrangle_{p_\bl}. \label{eqKLdef}
\end{equation}
This expression uses the result ${\frac{\partial Z(\boldsymbol{\lambda})^{-1}}{\partial \bl} = Z(\boldsymbol{\lambda})^{-1} \llangle\beta \frac{\partial H(\mathbf{q}; \bl)}{\partial \bl} \rrangle_{p_\bl} }$ but makes no assumption on the form of $p^*$.  Hence, an update rule for the $n^\text{th}$ update of $\bl^n$ is
\begin{equation}
	\bl^{n+1} \leftarrow \bl^n - \eta \left(\llangle\beta \frac{\partial H(\mathbf{q}; \bl^n)}{\partial \bl} \rrangle_{p^*} - \llangle\beta \frac{\partial H (\mathbf{q}; \bl^n)}{\partial \bl } \rrangle_{p_{\bl^n}} \right)
\end{equation}
where $\eta$ is a scalar learning rate, and the notation $\frac{H(\mathbf{q}; \bl^n)}{\partial \bl}$ indicates the gradient $\frac{H(\mathbf{q}; \bl)}{\partial \bl}$ evaluated at $\bl = \bl^n$ and averaged over the distribution $p^*(\mathbf{q})$.  Update rules of this kind are often used to train machine learning models such as restricted Boltzmann machines, where a practical issue is that of sufficiently sampling over the distributions $p^*$ and $p_{\bl^n}$.  Techniques such as CD-n have been proposed to efficiently perform this sampling \cite{carreira2005contrastive}; we do not concern ourselves with these issues here and assume that necessary averages over these distributions are accessible.  

We consider how to learn a target \textit{trajectory} $p^*(\mathbf{q},t)$, rather than a single static distribution $p^*(\mathbf{q})$.  Let us first assume that the target trajectory results from a driving protocol $\bl^*(t)$ which is quasi-static, so that 
\begin{equation}
	p^*(\mathbf{q},t) = p^\text{eq}_{\bl^*(t)}(\mathbf{q}) = Z(\boldsymbol{\lambda}^*(t))^{-1} e^{-\beta H(\mathbf{q}; \bl^*(t))}.
\end{equation} 
Considering quasi-static trial protocols $\bl^{i}(t)$, then one could use standard contrastive learning to learn this trajectory by applying updates to $\bl^{i}(t)$ of the form
\begin{equation}
	\bl^{n+1}(t) \leftarrow \bl^{n}(t) - \eta\Delta^\text{eq}
\end{equation}
where
\begin{eqnarray}
	\Delta^\text{eq} &= &  \frac{ \partial \mathcal{D}^\text{eq}}{\partial \bl } \equiv \frac{ \partial \mathcal{D}[p^\text{eq}_{\bl^*(t)}||p^\text{eq}_{\bl^{n}(t)}]}{\partial \bl } \nonumber \\
	&=& \llangle \beta \frac{\partial H(\mathbf{q}; \bl^{n}(t))}{\partial \bl } \rrangle_{p^\text{eq}_{\bl^*(t)}} - \llangle \beta \frac{H(\mathbf{q}; \bl^{n}(t))}{\partial \bl} \rrangle_{p^\text{eq}_{\bl^{n}(t)} } . \label{eqUpEq}
\end{eqnarray}
For these updates, we sample the gradients $\partial_\bl H(\mathbf{q}; \bl)$ of the Hamiltonian function in each distribution.   
The assumption of quasi-staticity allows one to break the dynamical problem into a set of \textit{independent} problems to which standard contrastive learning based on equilibrium distributions can be applied.  In SI Section \ref{secSIpath}, we illustrate this further using as an alternative starting cost function the KL divergence evaluated over path probabilities.

If the driving protocol $\bl^*(t)$ is not quasi-static then in principle we should not consider quasi-static trial protocols.  Away from equilibrium one can still formally write $p_\bl(\mathbf{q})$ as 
\begin{equation}
	p_{\bl}(\mathbf{q}) = \tilde{Z}(\bl)^{-1} e^{-\beta \tilde{H}(\mathbf{q}; \bl)} \label{eqFormal}
\end{equation}
where 
\begin{equation}
	\tilde{Z}(\boldsymbol{\lambda}) = \int d\mathbf{q}^{-\beta \tilde{H}(\mathbf{q}; \boldsymbol{\lambda})}.
\end{equation}
The function exponential weight $\tilde{H}$ is no longer the Hamiltonian function when the system is out of equilibrium. If we knew $\tilde{H}$, then to minimize Equation \ref{eqKLdef} we could sample the gradients $\partial_\bl \tilde{H}(\mathbf{q}; \bl)$ in updates of the form 
\begin{eqnarray}
	\Delta^\text{neq} &=&  \frac{\partial \mathcal{D}^\text{neq}}{\partial \bl } \equiv \frac{\partial  \mathcal{D}[p_{\bl^*(t)}||p_{\bl^{n}(t)}] }{\partial \bl } \nonumber \\
	&=& \llangle \beta \frac{\partial \tilde{H}(\mathbf{q}; \bl^n(t))}{\partial \bl } \rrangle_{p_{\bl^*(t)}} - \llangle \beta \frac{\partial \tilde{H}(\mathbf{q}; \bl^n(t))}{\partial \bl} \rrangle_{p_{\bl^{n}(t)} } .
\end{eqnarray}
However, $\tilde{H}$ and its gradients are not known in general for non-equilibrium processes, so this update cannot feasibly be implemented.  Instead, we consider the approximation to this update in which $\tilde{H}$ is replaced by the Hamiltonian $H$:
\begin{equation}
	\Delta^\text{app} \equiv \llangle \beta \frac{\partial H(\mathbf{q}; \bl^n(t))}{\partial\bl} \rrangle_{p_{\bl^*(t)}} - \llangle \beta \frac{\partial H (\mathbf{q}; \bl^n(t))}{\partial \bl} \rrangle_{p_{\bl^{n}(t)} },
\end{equation}
In contrast to $\Delta^\text{eq}$ the samples are taken with respect to the non-equilibrium distributions $p_{\bl^*(t)}$ and $p_{\bl^{n}(t)}$ rather than their quasi-static counterparts, and in contrast to $\Delta^\text{app}$ the samples are of gradients of $H$ rather than $\tilde{H}$.

\begin{figure}[h!]
	\begin{center}
		\includegraphics[width=0.6 \textwidth]{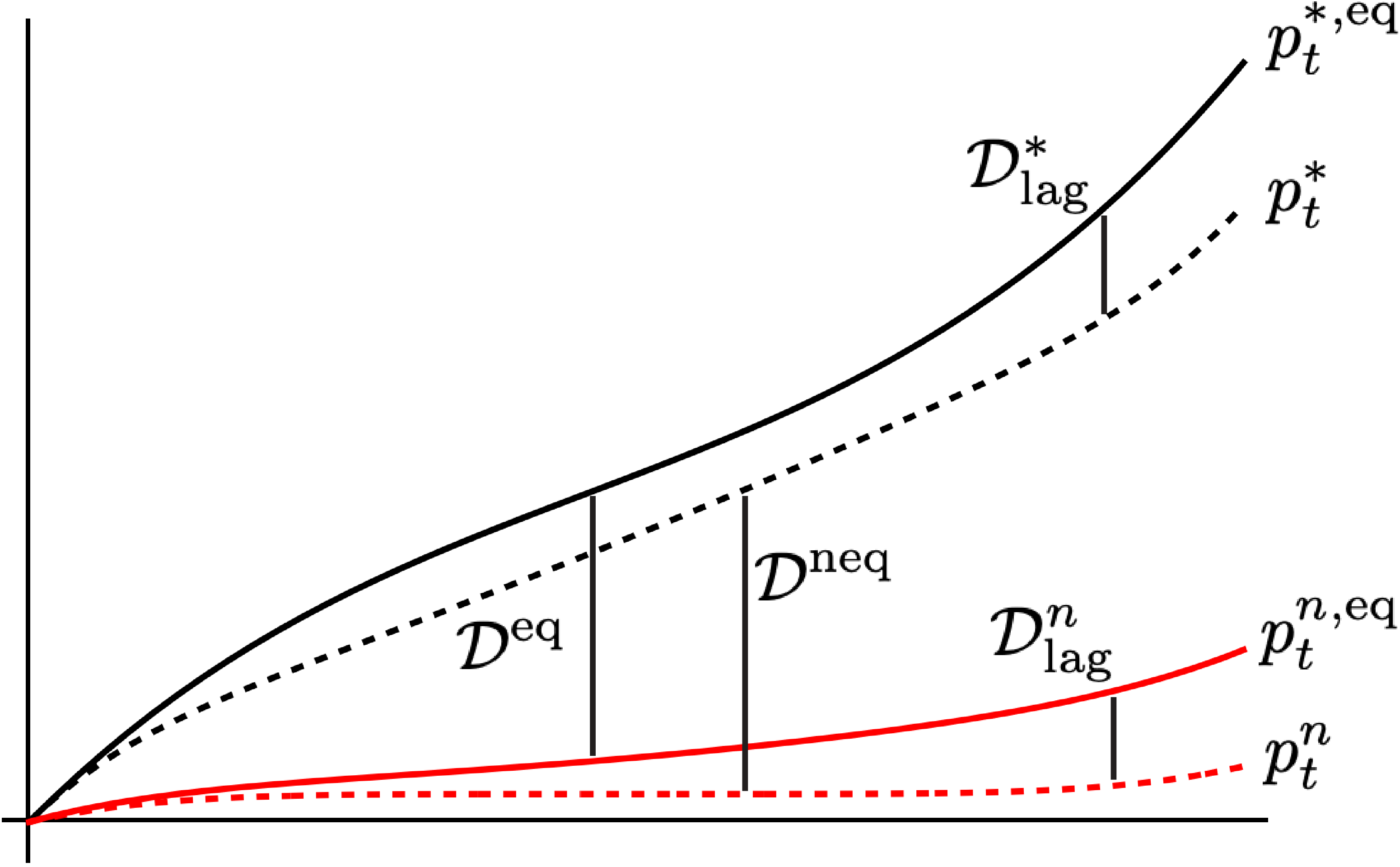}
		\caption{D) Schematic illustration of learning a non-equilibrium probability trajectory $p^*_t$ which lags behind a quasi-static version $p^{*,\text{eq}}_t$.  Similarly the trial trajectory $p^n_t$ lags behind a quasi-static version $p^{n,\text{eq}}_t$.  KL divergences can be defined between various pairs of these distributions as dicussed in the main text.  }
		\label{DissLagPlot}
	\end{center}
\end{figure}

To motivate why this approximate learning rule might work, we consider the view of non-autonomous dynamics described in Ref.\ \citenum{vaikuntanathan2009dissipation} as comprising an inevitable lag which develops between a non-equilibrium distribution and its quasi-static counterpart (in which the $\bl(t)$ is moved along the same geometrical path at an infinitely slow rate).  Schematically, this is depicted as the pair of black lines in SI Figure \ref{DissLagPlot}.  The KL divergence $\mathcal{D}^*_\text{lag} \equiv \mathcal{D}[p^*(t)||p^{*,\text{eq}}(t)]$ is in fact bounded by amount of work dissipated up to time $t$.  Similarly, there is a non-zero lag-related KL divergence $\mathcal{D}^n_\text{lag} \equiv \mathcal{D}[p^n(t)||p^{n,\text{eq}}(t)]$ which applies for the $n^\text{th}$ iteration of our trial process.  To close the distance between $p^n(t)$ and $p^*(t)$, we would ideally minimize $\mathcal{D}^\text{neq}$, which differs from $\mathcal{D}^\text{eq}$ due to the two lag-related divergences $\mathcal{D}^*_\text{lag}$ and $\mathcal{D}^n_\text{lag}$.
$\mathcal{D}^\text{neq}$ and its gradient $\Delta^\text{neq}$ are unfortunately not known, but flow down the $\Delta^\text{neq}$ will have the same fixed point as flow down $\Delta^\text{eq}$ or $\Delta^\text{app}$.  This shared fixed point will occur at $\bl^n(t') = \bl^*(t')$ for $t'\leq t$.  Among these three gradients, only $\Delta^\text{app}$ can be computed in general as it involves samples over the accessible distributions $p^*(t)$ and $p^n(t)$, (rather than their unknown quasi-static counterparts) and involves the known model Hamiltonian $H$, rather than the unknown exponential weight $\tilde{H}$.  Even though the vector $\Delta^\text{app}$ will systematically differ from $\Delta^\text{neq}$, as long as there is a positive projection between the two then $\Delta^\text{app}$ will remain a descent direction and will converge to the correct minimum (see Figure \ref{Fig1Schematic}C of the main text). 

\subsection{Path integral formulation}\label{secSIpath}
An alternative approach is to consider minimizing over the whole trajectory $\bl(t)$ using a path-integral formulation of the KL divergence
\begin{equation}
	D_\text{path}\left[\mathcal{P}^*[\mathbf{q}(t)] || \mathcal{P}_{\bl^n(t)}[\mathbf{q}(t)]\right] \equiv \int \mathcal{D}[\mathbf{q}(t)]\mathcal{P}^*[\mathbf{q}(t)]\ln \frac{\mathcal{P}^*[\mathbf{q}(t)]}{\mathcal{P}_{\bl^n(t)}[\mathbf{q}(t)]}.
\end{equation}
We can formally write the path probabilities under a protocol $\bl(t)$ as 
\begin{equation}
	\mathcal{P}_{\bl(t)}[\mathbf{q}(t)] = \mathcal{Z}[\bl(t)]^{-1}e^{-\beta \mathcal{A}[\mathbf{q}(t); \bl(t)]}
\end{equation}
for some action integral $\mathcal{A}[\mathbf{q}(t); \bl(t)]$ and normalization 
\begin{equation}
	\mathcal{Z}[\bl(t)] = \int \mathcal{D}[\mathbf{q}(t)]e^{-\beta \mathcal{A}[\mathbf{q}(t); \bl(t)]}.
\end{equation}
The functional derivative of $D_\text{path}\left[\mathcal{P}^*[\mathbf{q}(t)] || \mathcal{P}_{\bl^n(t)}[\mathbf{q}(t)]\right]$ with respect to $\bl(t)$ is (cf. Equation \ref{eqKLdef})
\begin{equation}
	\frac{\delta D_\text{path}}{\delta \bl(t)} = \llangle \beta \frac{\delta \mathcal{A}}{\delta\bl(t)}\rrangle_{\mathcal{P}^*} - \llangle \beta \frac{\delta \mathcal{A}}{\delta\bl(t)} \rrangle_{\mathcal{P}_{\bl^n(t)}} \label{eqPathKL}
\end{equation}
where the averages are functional integrals using the path probabilities.  

In principle, if one knows the action describing the system's evolution and can evaluate samples over path distributions, then Equation \ref{eqPathKL} can be used to perform gradient descent in path space to optimize $\bl(t)$.  Typical actions, such as the Onsager-Machlup action, involve time derivatives of $\mathbf{q}$, which will cause Equation \ref{eqPathKL} to depend on these quantities as well \cite{onsager1953fluctuations}.  This approach is thus qualitatively different from optimizing at each time point independently using the non-equilibrium distribution potential $\widetilde{H}$ which depends only on $\mathbf{q}$, not its time derivatives.  However, we now show that for quasi-static protocols, Equation \ref{eqPathKL} implies the same increment as Equation \ref{eqUpEq}.

For a quasi-static case, the path probability can be written as
\begin{align}
	\mathcal{P}[\mathbf{q}(t)] &= \lim_{N_t\rightarrow \infty} p(\mathbf{q}_0) p(\mathbf{q}_1|\mathbf{q}_0)\cdots p(\mathbf{q}_{N}|\mathbf{q}_{N-1}) \nonumber \\ 
	&= p(\mathbf{q}_0) p(\mathbf{q}_1)\cdots p(\mathbf{q}_{N})
\end{align}
where we have discretized the trajectory of length $T$ into $N_t$ intervals.  The second line follows because for quasi-static systems the system lacks memory and the conditional probabilities simplify.  Each distribution $p(\mathbf{q}_i)$ is given by the Boltzmann distribution under the current value of the work parameter, so that
\begin{align}
	\mathcal{P}[\mathbf{q}(t)] =& \lim_{N_t \rightarrow \infty} \left(\prod_{n=0}^{N_t}Z(\bl_n) \right)^{-1}\exp \left(-\beta \sum_{n=0}^{N_t} H(\mathbf{q}_n ; \bl_n) \right) \nonumber \\
	=& \mathcal{Z}^{\text{eq}}[\bl(t)]^{-1} \exp \left( -\beta \frac{1}{T} \int_0^TH(\mathbf{q}(t) ; \bl(t)) dt \right),
\end{align}
where 
\begin{equation}
	\mathcal{Z}^{\text{eq}}[\bl(t)] = \int \mathcal{D}[\mathbf{q}(t)]\exp \left( -\beta \frac{1}{T} \int_0^TH(\mathbf{q}(t) ; \bl(t)) dt \right).
\end{equation}
This result implies that for a quasi-static process, $\mathcal{A}[\mathbf{q}(t); \bl(t)] = \int_0^TH(\mathbf{q}(t) ; \bl(t)) dt$.  Evaluating Equation \ref{eqPathKL} then gives
\begin{equation}
	\frac{\delta D_\text{path}}{\delta \bl(t)} = \llangle \beta \frac{H(\mathbf{q}(t); \bl(t))}{\partial \bl} \rrangle_{\mathcal{P}^*} - \llangle \beta \frac{H(\mathbf{q}(t); \bl(t))}{\partial \bl} \rrangle_{\mathcal{P}^n_{\bl(t)}}.
\end{equation}
This resulting derivative is local in time and can be equivalently expressed as in Equation \ref{eqUpEq}.  This result makes precise the statement that quasi-staticity breaks up the global trajectory-level problem into a manifold of independent temporally local problems.

\subsection{Systems under linear response}
For systems under weak driving we can use adiabatic perturbation to find expressions for the how the approximate gradient $\Delta^\text{app}$ differs from correct gradients $\Delta^\text{neq}$ and $\Delta^\text{eq}$.  The non-equilibrium distribution in this regime can be written as
\begin{equation}
	p_{\bl(t)}(\mathbf{q}) = p^\text{eq}_{\bl(t)}(\mathbf{q}) + \epsilon p^1_{\bl(t)}(\mathbf{q}) \label{eqlinresp1}
\end{equation}
where the superscript $1$ indicates the first order correction to the quasi-static time-dependent distribution $p^\text{eq}_{\bl(t)}$.  This correction can be expressed in terms of the inverse of the dynamical evolution operator and the time derivative of the protocol $\bl(t)$, but we will not use this expression here \cite{parrondo1998reversible, funo2020shortcuts}.  Due to normalization of $p^\text{eq}_{\bl(t)}$ and $p_{\bl(t)}$, the integral over $\mathbf{q}$ of $p^1_{\bl(t)}$ is zero.

We first rewrite Equation \ref{eqlinresp1} as 
\begin{eqnarray}
	p_{\bl(t)}(\mathbf{q}) &=& p^\text{eq}_{\bl(t)}(\mathbf{q}) \left(1 + \epsilon \frac{p^1_{\bl(t)}(\mathbf{q})}{p^\text{eq}_{\bl(t)}(\mathbf{q})} \right) \nonumber \\
	&=& Z(\bl({t}))^{-1} e^{-\beta H(\mathbf{q};\bl(t)) + m(\mathbf{q};\bl(t))} \nonumber \\
	&\equiv&Z(\bl({t}))^{-1} e^{-\beta \tilde{H}(\mathbf{q};\bl(t))} \label{eqlinresp2}
\end{eqnarray}
where 
\begin{equation}
	m(\mathbf{q};\bl(t)) \equiv \ln \left(1 + \epsilon \frac{p^1_{\bl(t)}(\mathbf{q})}{p^\text{eq}_{\bl(t)}(\mathbf{q})}\right).
\end{equation}
To confirm that $Z(\bl(t))$ is the proper normalization factor for both the exponential weights $-\beta H(\mathbf{q};\bl(t))$ and $-\beta \tilde{H}(\mathbf{q};\bl(t))$, we evaluate the integral
\begin{eqnarray}
	\int d\mathbf{q} e^{-\beta \tilde{H}(\mathbf{q};\bl(t))} &=& \int d\mathbf{q} \left(1 + \frac{p^1_{\bl(t)}(\mathbf{q})}{p^\text{eq}_{\bl(t)}(\mathbf{q})}\right) e^{-\beta H(\mathbf{q};\bl(t))} \nonumber \\ 
	&=& Z(\bl(t)) + Z(\bl(t))\int d\mathbf{q} p^1_{\bl(t)}(\mathbf{q}) \nonumber \\ 
	&=& Z(\bl(t)).
\end{eqnarray}
Thus, Equation \ref{eqlinresp2} is the correct non-equilibrium distribution in the form of Equation \ref{eqFormal} above.  

We next evaluate the gradients $\Delta^\text{eq}$, $\Delta^\text{app}$, and $\Delta^\text{neq}$.  We have
\begin{equation}
	\Delta^\text{eq} \equiv \llangle \beta  \frac{\partial H(\mathbf{q}; \bl^n(t))}{\partial \bl}\rrangle_{p^\text{eq}_{\bl^*(t)}} - \llangle \beta  \frac{\partial H(\mathbf{q}; \bl^n(t))}{\partial \bl}\rrangle_{p^\text{eq}_{\bl^n(t)}}
\end{equation}
which corresponds to the limit $\epsilon \rightarrow 0$.  The gradient $\Delta^\text{app}$ is 
\begin{eqnarray}
	\Delta^\text{app} &\equiv& \llangle \beta  \frac{\partial H(\mathbf{q}; \bl^n(t))}{\partial \bl}\rrangle_{p_{\bl^*(t)}} - \llangle \beta  \frac{\partial H(\mathbf{q}; \bl^n(t))}{\partial \bl}\rrangle_{p_{\bl^n(t)}} \nonumber \\
	&=& \Delta^\text{eq} + \epsilon \left(\llangle \beta  \frac{\partial H(\mathbf{q}; \bl^n(t))}{\partial \bl}\rrangle_{p^1_{\bl^*(t)}} - \llangle \beta  \frac{\partial H(\mathbf{q}; \bl^n(t))}{\partial \bl}\rrangle_{p^1_{\bl^n(t)}} \right). \label{eqdapplinresp}
\end{eqnarray}
If we set $p^1_{\bl^*(t)} = p^1_{\bl^n(t)}$ then the correction to $\Delta^\text{app} = \Delta^\text{eq}$ vanishes.  We thus see that the difference between $\Delta^\text{app}$ and $\Delta^\text{eq}$ depends on the degree to which ``non-equilibrium components'' of the distributions $p_{\bl^*(t)}$ and $p_{\bl^n(t)}$ differ.  

To evaluate $\Delta^\text{neq}$, we first need to find (expanding the logarithm for small $\epsilon$)
\begin{eqnarray}
	\llangle \frac{\partial m(\mathbf{q};\bl)}{\partial \bl} \rrangle_{p_{\bl(t)}} &=& \epsilon \llangle  \frac{\partial}{\partial \bl} \frac{p^1_{\bl}(\mathbf{q})}{p^\text{eq}_{\bl}(\mathbf{q})} \rrangle_{p_{\bl(t)}} + \mathcal{O}(\epsilon^2).
\end{eqnarray}
The order $\epsilon$ term can be expressed as (remembering that integrals over $p_{\bl}^1(\mathbf{q})$ are zero)
\begin{eqnarray}
	\epsilon \int d\mathbf{q}\left( \frac{\partial}{\partial \bl} \frac{p^1_{\bl}(\mathbf{q})}{p^\text{eq}_{\bl}(\mathbf{q})}\right) \left(p^\text{eq}_{\bl}(\mathbf{q}) + \epsilon p^1_{\bl}(\mathbf{q})\right)
	&=& \epsilon \int d\mathbf{q}\left( \frac{\partial}{\partial \bl} \frac{p^1_{\bl}(\mathbf{q})}{p^\text{eq}_{\bl}(\mathbf{q})}\right) p^\text{eq}_{\bl}(\mathbf{q}) + \mathcal{O}(\epsilon^2) \nonumber \\
	&=& \epsilon \llangle  \frac{\partial}{\partial \bl} \frac{p^1_{\bl}(\mathbf{q})}{p^\text{eq}_{\bl}(\mathbf{q})} \rrangle_{p^\text{eq}_{\bl(t)}} + \mathcal{O}(\epsilon^2) \label{eqSItemp} .
\end{eqnarray}
We can alternatively write the order $\epsilon$ term as
\begin{eqnarray}
	\epsilon \llangle  \frac{\partial}{\partial \bl} \frac{p^1_{\bl}(\mathbf{q})}{p^\text{eq}_{\bl}(\mathbf{q})} \rrangle_{p^\text{eq}_{\bl(t)}} &=& - \epsilon \int  d\mathbf{q}p^1_{\bl}(\mathbf{q}) \frac{ \partial \ln p^\text{eq}_{\bl}(\mathbf{q})}{\partial \bl} \label{eqSItemp2}  \nonumber \\
	&=&  \epsilon \int  d\mathbf{q}  p^1_{\bl}(\mathbf{q}) \beta \frac{ \partial H(\mathbf{q};\bl)}{\partial \bl} \nonumber \\
	&=& \epsilon \llangle \beta \frac{\partial H(\mathbf{q};\bl)}{\partial \bl} \rrangle_{p^1_{\bl}}. \label{eqlinrespdiff}
\end{eqnarray}
With this, we can write
\begin{eqnarray}
	\Delta^\text{neq} &\equiv& \llangle \beta  \frac{\partial \tilde{H}(\mathbf{q}; \bl^n(t))}{\partial \bl}\rrangle_{p_{\bl^*(t)}} - \llangle \beta  \frac{\partial \tilde{H}(\mathbf{q}; \bl^n(t))}{\partial \bl}\rrangle_{p_{\bl^n(t)}}  \nonumber \\
	&\equiv&  \llangle \beta  \frac{\partial H(\mathbf{q}; \bl^n(t))}{\partial \bl}\rrangle_{p_{\bl^*(t)}} - \llangle \beta  \frac{\partial H(\mathbf{q}; \bl^n(t))}{\partial \bl}\rrangle_{p_{\bl^n(t)}} - \left(\llangle  \frac{\partial m(\mathbf{q}; \bl^n(t))}{\partial \bl}\rrangle_{p_{\bl^*(t)}} - \llangle \beta  \frac{\partial m(\mathbf{q}; \bl^n(t))}{\partial \bl}\rrangle_{p_{\bl^n(t)}}\right) \nonumber \\
	&=& \Delta^\text{app} - \epsilon \left(\llangle \frac{\partial }{\partial \bl}\frac{p^1_{\bl^n(t)}}{p^\text{eq}_{\bl^n(t)}} \rrangle_{p^\text{eq}_{\bl^*(t)}} - \llangle \beta \frac{\partial H(\mathbf{q};\bl^n(t))}{\partial \bl} \rrangle_{p^1_{\bl^n(t)}} \right) \nonumber + \mathcal{O}(\epsilon^2). \label{eqSItemp3}
\end{eqnarray}
We note that the the term $\llangle  \frac{ \partial m(\mathbf{q}; \bl^n(t))}{\partial \bl} \rrangle_{p_{\bl^*(t)}}$ does not simplify like $\llangle  \frac{m(\mathbf{q}; \bl^n(t))}{\partial \bl}\rrangle_{p_{\bl^n(t)}}$ does, because the average is taken over $p_{\bl^*(t)}$ while the integrand is evaluated for $p_{\bl^n(t)}$, which prevents passing from Equation \ref{eqSItemp} to \ref{eqSItemp2}.  We can rewrite the term
\begin{equation}
	\llangle \frac{\partial }{\partial \bl}\frac{p^1_{\bl^n(t)}}{p^\text{eq}_{\bl^n(t)}} \rrangle_{p^\text{eq}_{\bl^*(t)}} = \int d\mathbf{q} \frac{p^\text{eq}_{\bl^*(t)}(\mathbf{q})}{p^\text{eq}_{\bl^n(t)}(\mathbf{q})}\frac{\partial}{\partial \bl} p^1_{\bl^n(t)}(\mathbf{q}) +  \int d\mathbf{q} \left( \frac{p^\text{eq}_{\bl^*(t)}(\mathbf{q})}{p^\text{eq}_{\bl^n(t)}(\mathbf{q})}\right) p^1_{\bl^n(t)}(\mathbf{q}) \left(\frac{\partial}{\partial \bl} \ln\left(Z(\bl^n(t))\right) + \beta\frac{\partial}{\partial \bl} H(\mathbf{q}; \bl^n(t))  \right). \label{eqfraclinresp}
\end{equation}
If we set $p^\text{eq}_{\bl^*(t)} = p^\text{eq}_{\bl^n(t)}$, Equation \ref{eqfraclinresp} simplifies as
\begin{eqnarray}
	\llangle \frac{\partial }{\partial \bl}\frac{p^1_{\bl^n(t)}}{p^\text{eq}_{\bl^n(t)}} \rrangle_{p^\text{eq}_{\bl^*(t)}} &=& \frac{\partial}{\partial \bl} \int d\mathbf{q} p^1_{\bl^n(t)}(\mathbf{q}) +  \left(\frac{\partial}{\partial \bl} \ln\left(Z(\bl^n(t))\right)\right) \int d\mathbf{q} p^1_{\bl^n(t)}(\mathbf{q})  + \int d\mathbf{q}p^1_{\bl^n(t)}(\mathbf{q})  \beta\frac{\partial}{\partial \bl} H(\mathbf{q}; \bl^n(t))  \nonumber \\
	&=& \llangle \beta \frac{\partial H(\mathbf{q};\bl^n(t))}{\partial \bl} \rrangle_{p^1_{\bl^n(t)}}
\end{eqnarray}
which cancels the term $-\llangle \beta \frac{\partial H(\mathbf{q};\bl^n(t))}{\partial \bl} \rrangle_{p^1_{\bl^n(t)}}$ in the order $\epsilon$ correction of $\Delta^\text{app}$ to $\Delta^\text{neq}$ in Equation \ref{eqSItemp3}.  We thus see that the difference between $\Delta^\text{app}$ and $\Delta^\text{neq}$ depends on the degree to which the ``equilibrium components'' $p^\text{eq}_{\bl^*(t)}$ and $p^\text{eq}_{\bl^n(t)}$ differ from each other.   

To go further and demonstrate convergence of the approximate update $\Delta^\text{app}$, we first recall a result from Refs.\ \citenum{sivak2012thermodynamic} and \citenum{zulkowski2012geometry} that for a system under linear response 
\begin{equation}
	\llangle X_\alpha(\bl(t)) \rrangle_{p_{\bl(t)}} - \llangle X_\alpha(\bl(t)) \rrangle_{p^\text{eq}_{\bl(t)}} \approx \zeta_{\alpha \gamma}(\bl(t)) \frac{d \lambda_\gamma(t)}{dt} \label{eqlinrespSC1}
\end{equation}
where 
\begin{equation}
	X_\alpha(\bl(t)) \equiv -\frac{\partial H(\mathbf{q}; \bl(t))}{\partial \lambda_\alpha}
\end{equation}
is the thermodynamic force conjugate to the parameter component $\lambda_\alpha$, and we use the shorthand notation $\frac{d\lambda_\gamma(t)}{dt} = \frac{d\lambda_\gamma}{dt}|_{\lambda_\gamma = \lambda_\gamma(t)}$.  The time integrated matrix 
\begin{equation}
	\zeta_{\alpha\gamma}(\bl) = \beta \int_0^\infty dt''\llangle \Delta X_\alpha(\bl, t'=0) \Delta X_\gamma (\bl,t'=t'') \rrangle_{p^\text{eq}_{\bl}}
\end{equation}
measures the covariance between the stochastic quantities $\Delta X_\alpha(\bl, t'=0)$, the fluctuations in $X_\alpha$ under parameter values $\bl$ measured at time $t'=0$, and $\Delta X_\gamma (\bl,t'=t'')$. The matrix $\zeta_{\alpha\gamma}(\bl)$ is also the Kirkwood expression of the friction tensor \cite{kirkwood1946statistical}.  Writing the difference between the gradients $\Delta^\text{app}$ and $\Delta^\text{eq}$ as 
\begin{eqnarray}
	\Delta^\text{app} - \Delta^\text{eq} = -\beta\left(\left(\llangle \mathbf{X}(\bl^n(t))\rrangle_{p_{\bl^*(t)}} - \llangle \mathbf{X}(\bl^n(t))\rrangle_{p^\text{eq}_{\bl^*(t)}} \right) - \left(\llangle \mathbf{X}(\bl^n(t))\rrangle_{p_{\bl^n(t)}} - \llangle \mathbf{X}(\bl^n(t))\rrangle_{p^\text{eq}_{\bl^n(t)}} \right)\right), \label{eqlinrespgradddiff}
\end{eqnarray}
the final term can be written as
\begin{equation}
	\llangle \mathbf{X}(\bl^n(t))\rrangle_{p_{\bl^n(t)}} - \llangle \mathbf{X}(\bl^n(t))\rrangle_{p^\text{eq}_{\bl^n(t)}} = \xi_{\alpha \gamma}(\bl^n(t))\frac{d\lambda^n(t)}{dt}.
\end{equation}
In the vicinity of convergence, when $\bl^n(t)$ nears $\bl^*(t)$, we can write
\begin{eqnarray}
	\frac{\partial H(\mathbf{q};\bl^n(t))}{\partial \lambda_\alpha} = \frac{\partial H(\mathbf{q};\bl^*(t))}{\partial \lambda_\alpha} + \left(\lambda_\gamma^n(t) - \lambda_\gamma^*(t)\right)\frac{\partial^2 H(\mathbf{q};\bl^*(t))}{\partial \lambda_\gamma \partial 
		\lambda_\alpha} \nonumber \\
	\equiv \frac{\partial H(\mathbf{q};\bl^*(t))}{\partial \lambda_\alpha} + \delta\lambda_\gamma(t) \Delta_{\gamma \alpha}H(\mathbf{q}; \bl^*(t)).
\end{eqnarray}
With this we can write the first term on on the right of Equation \ref{eqlinrespgradddiff} as
\begin{eqnarray}
	\llangle X_\alpha(\bl^n(t))\rrangle_{p_{\bl^*(t)}} - \llangle X_\alpha (\bl^n(t))\rrangle_{p^\text{eq}_{\bl^*(t)}} &=& \llangle X_\alpha(\bl^*(t))\rrangle_{p_{\bl^*(t)}} - \llangle X_\alpha(\bl^*(t))\rrangle_{p^\text{eq}_{\bl^*(t)}} \nonumber \\
	&& - \delta\lambda_\gamma(t)\left(\llangle \Delta_{\gamma \alpha}H(\mathbf{q}; \bl^*(t)) \rrangle_{p_{\bl^*(t)}} - \llangle \Delta_{\gamma \alpha}H(\mathbf{q}; \bl^*(t)) \rrangle_{p^\text{eq}_{\bl^*(t)}} \right) \nonumber \\
	&=& \xi_{\alpha \gamma}(\bl^*(t))\frac{d\lambda^*_\gamma}{dt} \nonumber \\
	&& - \delta\lambda_\gamma(t)\left(\llangle \Delta_{\gamma \alpha}H(\mathbf{q}; \bl^*(t)) \rrangle_{p_{\bl^*(t)}} - \llangle \Delta_{\gamma \alpha}H(\mathbf{q}; \bl^*(t)) \rrangle_{p^\text{eq}_{\bl^*(t)}} \right).
\end{eqnarray}
We also have that
\begin{eqnarray}
	\Delta^\text{eq}_\alpha &=& -\beta\left( \llangle X_\alpha(\bl^n(t)) \rrangle_{p^\text{eq}_{\bl^*(t)}} - \llangle X_\alpha(\bl^n(t)) \rrangle_{p^\text{eq}_{\bl^n(t)}} \right) \nonumber \\
	&=& -\beta \left( \llangle X_\alpha(\bl^*(t)) \rrangle_{p^\text{eq}_{\bl^*(t)}} - \llangle X_\alpha(\bl^n(t)) \rrangle_{p^\text{eq}_{\bl^n(t)}} - \delta\lambda_\gamma(t) \llangle \Delta_{\gamma \alpha}H(\mathbf{q}; \bl^*(t))\rrangle_{p^\text{eq}_{\bl^*(t)}} \right). 
\end{eqnarray}
The terms $\llangle X_\alpha(\bl(t)) \rrangle_{p^\text{eq}_{\bl(t)}}$ can be expressed in terms of the free energy $F(\bl) = - \beta^{-1} \ln Z(\bl)$ as 
\begin{equation}
	\llangle X_\alpha(\bl(t)) \rrangle_{p^\text{eq}_{\bl(t)}} = - \frac{\partial F(\bl(t))}{\partial \lambda_\alpha},
\end{equation}
so that
\begin{equation}
	\Delta^\text{eq}_\alpha = -\beta \left(\frac{\partial F(\bl^n(t))}{\partial \lambda_\alpha} - \frac{\partial F(\bl^*(t))}{\partial \lambda_\alpha} - \delta\lambda_\gamma(t) \llangle \Delta_{\gamma \alpha}H(\mathbf{q}; \bl^*(t))\rrangle_{p^\text{eq}_{\bl^*(t)}} \right).
\end{equation}
The free energy $F(\bl^n(t))$ can be expanded as 
\begin{equation}
	F(\bl^n(t)) = F(\bl^*(t)) + \delta \lambda_\gamma(t) \frac{\partial F(\bl^*(t))}{\partial \lambda_\gamma}
\end{equation}
so that 
\begin{equation}
	\frac{\partial F(\bl^n(t))}{\partial \lambda_\alpha} - \frac{\partial F(\bl^*(t))}{\partial \lambda_\alpha} = \delta \lambda_\gamma(t) \frac{\partial^2 F(\bl^*(t))}{\partial \lambda_\gamma\partial \lambda_\alpha} \equiv \delta \lambda_\gamma(t) \Delta_{\gamma \alpha}F(\bl^*(t)).
\end{equation}
Putting everything together, we can rewrite Equation \ref{eqlinrespgradddiff} as
\begin{eqnarray}
	\Delta^\text{app}_\alpha &=& \beta\bigg( -\delta \lambda_\gamma(t) \Delta_{\gamma \alpha}F(\bl^*(t)) + \delta\lambda_\gamma(t) \llangle \Delta_{\gamma \alpha}H(\mathbf{q}; \bl^*(t))\rrangle_{p^\text{eq}_{\bl^*(t)}} \nonumber \\
	&& -  \xi_{\alpha \gamma}(\bl^*(t))\frac{d\lambda^*_\gamma}{dt} + \delta\lambda_\gamma(t)\left(\llangle \Delta_{\gamma \alpha}H(\mathbf{q}; \bl^*(t)) \rrangle_{p_{\bl^*(t)}} - \llangle \Delta_{\gamma \alpha}H(\mathbf{q}; \bl^*(t)) \rrangle_{p^\text{eq}_{\bl^*(t)}} \right)
	+ \xi_{\alpha \gamma}(\bl^n(t))\frac{d\lambda^n(t)}{dt}
	\bigg) \nonumber \\
	&=& \beta\bigg(\xi_{\alpha \gamma}(\bl^n(t))\frac{d\lambda^n(t)}{dt} - \xi_{\alpha \gamma}(\bl^*(t))\frac{d\lambda^*_\gamma}{dt} + \delta\lambda_\gamma(t)\llangle \Delta_{\gamma \alpha}H(\mathbf{q}; \bl^*(t)) \rrangle_{p_{\bl^*(t)}} - \delta \lambda_\gamma(t) \Delta_{\gamma \alpha}F(\bl^*(t)) \bigg).
\end{eqnarray}
Introducing $\delta \xi_{\alpha \gamma}(t) = \xi_{\alpha \gamma}(\bl^n(t)) - \xi_{\alpha \gamma}(\bl^*(t))$, we rewrite this as 
\begin{equation}
	\Delta^\text{app}_\alpha = \beta\bigg( \xi_{\alpha \gamma}^n \frac{d \delta \lambda_\gamma(t)}{dt} + \delta \xi_{\alpha \gamma}\frac{d \lambda_\gamma^*}{dt}+ \delta\lambda_\gamma(t)\left( \llangle \Delta_{\gamma \alpha}H(\mathbf{q}; \bl^*(t)) \rrangle_{p_{\bl^*(t)}} - \Delta_{\gamma \alpha}F(\bl^*(t))\right)\bigg). \label{eqlinresppenform}
\end{equation}
We can further simplify the last term of this expression by first writing
\begin{equation}
	\Delta_{\gamma \alpha}F(\bl^*(t)) =
	\frac{\partial}{\partial \lambda_\gamma} \int d \mathbf{q} p^\text{eq}_{\bl^*(t)}\frac{\partial H(\mathbf{q};\bl^*(t))}{\partial \lambda_\alpha} = \int d \mathbf{q} \frac{\partial}{\partial \lambda_\gamma} p^\text{eq}_{\bl^*(t)}\frac{\partial H(\mathbf{q};\bl^*(t))}{\partial \lambda_\alpha} + \llangle \Delta_{\gamma \alpha}H(\mathbf{q}; \bl^*(t)) \rrangle_{p^\text{eq}_{\bl^*(t)}}.
\end{equation}
Straightforward evaluation gives
\begin{equation}
	\frac{\partial}{\partial \lambda_\gamma} p^\text{eq}_{\bl^*(t)} = p^\text{eq}_{\bl^*(t)}\left(\llangle \beta\frac{\partial H(\mathbf{q}; \bl^*(t))}{\partial \lambda_\gamma} \rrangle_{ p^\text{eq}_{\bl^*(t)}} - \beta \frac{\partial H(\mathbf{q}; \bl^*(t))}{\partial \lambda_\gamma}\right)
\end{equation}
so that 
\begin{eqnarray}
	\int d \mathbf{q} \frac{\partial}{\partial \lambda_\gamma} p^\text{eq}_{\bl^*(t)}\frac{\partial H(\mathbf{q};\bl^*(t))}{\partial \lambda_\alpha} &=& - \beta \left( \llangle \frac{\partial H(\mathbf{q}; \bl^*(t))}{\partial \lambda_\gamma} \frac{\partial H(\mathbf{q}; \bl^*(t))}{\partial \lambda_\alpha} \rrangle_{ p^\text{eq}_{\bl^*(t)}}- \llangle \frac{\partial H(\mathbf{q}; \bl^*(t))}{\partial \lambda_\gamma} \rrangle_{ p^\text{eq}_{\bl^*(t)}}  \llangle \frac{\partial H(\mathbf{q}; \bl^*(t))}{\partial \lambda_\alpha} \rrangle_{ p^\text{eq}_{\bl^*(t)}} \right) \nonumber \\
	&\equiv& - \text{Cov}_{p^\text{eq}_{\bl^*(t)}}\left(\frac{\partial H(\mathbf{q}; \bl^*(t))}{\partial \lambda_\gamma} \frac{\partial H(\mathbf{q}; \bl^*(t))}{\partial \lambda_\alpha} \right)
\end{eqnarray}
Additionally using 
\begin{equation}
	\llangle \Delta_{\gamma \alpha}H(\mathbf{q}; \bl^*(t)) \rrangle_{p_{\bl^*(t)}} = \llangle \Delta_{\gamma \alpha}H(\mathbf{q}; \bl^*(t)) \rrangle_{p^\text{eq}_{\bl^*(t)}} + \epsilon \llangle \Delta_{\gamma \alpha}H(\mathbf{q}; \bl^*(t)) \rrangle_{p^1_{\bl^*(t)}}
\end{equation}
we can rewrite Equation \ref{eqlinresppenform}
\begin{equation}
	\Delta^\text{app}_\alpha = \beta\bigg( \xi_{\alpha \gamma}^n \frac{d \delta \lambda_\gamma(t)}{dt} + \delta \xi_{\alpha \gamma}\frac{d \lambda_\gamma^*}{dt}+ \delta\lambda_\gamma(t)\left(\text{Cov}_{p^\text{eq}_{\bl^*(t)}}\left(\frac{\partial H(\mathbf{q}; \bl^*(t))}{\partial \lambda_\gamma} \frac{\partial H(\mathbf{q}; \bl^*(t))}{\partial \lambda_\alpha} \right) +  \epsilon \llangle \Delta_{\gamma \alpha}H(\mathbf{q}; \bl^*(t)) \rrangle_{p^1_{\bl^*(t)}} \right)\bigg). \label{eqlinresppenapp}
\end{equation}
Considering that the quantities $\frac{d \lambda^*_\gamma}{dt}$ and $\epsilon$ are small due to the assumption of linear response, and that $\delta \lambda_\gamma(t)$ and $\delta \xi_{\alpha \gamma}$ are small due to the vicinity of convergence, only the term proportional to the covariance matrix survives to first order in Equation \ref{eqlinresppenapp}.  Hence, we have
\begin{equation}
	\Delta^\text{app}_\alpha \approx \beta g_{\alpha \gamma} \left(\lambda^n_\gamma(t) - \lambda^*_\gamma(t) \right)
\end{equation}
where 
\begin{equation}
	g_{\alpha \gamma} \equiv  \text{Cov}_{p^\text{eq}_{\bl^*(t)}}\left(\frac{\partial H(\mathbf{q}; \bl^*(t))}{\partial \lambda_\alpha}, \frac{\partial H(\mathbf{q}; \bl^*(t))}{\partial \lambda_\gamma} \right)
\end{equation}
is shown in Ref.\ \citenum{crooks2007measuring} to define a thermodynamic metric tensor which is equal to the Fisher information matrix.  Importantly, this tensor is a covariance matrix and hence positive semi-definite.  Flowing down $\Delta^\text{app}$ will therefore locally attract $\lambda^n_\alpha(t)$ toward the fixed point $\lambda^*_\alpha(t)$, as desired.

\subsection{Systems with linear forces}
Here we treat a class of non-autonomous systems in which linear forces act on the system degrees of freedom.  The overdamped stochastic dynamics for a vector $\mathbf{q}$ can be written as
\begin{equation}
	\partial_t q_i = - \mu K_{ij}\left(q_j - a_j \right) + \xi_i \label{eqLinDyn}
\end{equation}
where $\mu$ is a mobility, $K_{ij}(\bl_K(t)) = K_{ji}(\bl_K(t))$ is a symmetric coupling matrix parameterized by the protocol $\bl_K(t)$, $a_i(\bl_a(t))$ is a rest value parameterized by the protocol $\bl_a(t)$, and $\eta_i$ is an isotropic white noise obeying
\begin{equation}
	\llangle \xi_i(t) \xi_j(t') \rrangle = 2 \mu \beta^{-1} \delta_{ij} \delta(t-t').
\end{equation}
The corresponding Fokker-Planck equation is
\begin{equation}
	\partial_t p(\mathbf{q}) = \mu K_{ij}\partial_i\left((q_j - a_j) p(\mathbf{q}) \right) + D \partial_i \partial_i p(\mathbf{q})
\end{equation}
where $D = \mu \beta^{-1}$.  For a given value of the protocols $\bl \equiv (\bl_K, \ \bl_a)$, corresponding to $K_{ij}$ and $a_i$, there is an equilibrium probability distribution over $\mathbf{q}$ given by the multidimensional Gaussian
\begin{eqnarray}
	p_\bl^\text{eq}(\mathbf{q}) &=& \left(\frac{\beta \text{det}\left(\mathbf{K}\right)}{(2\pi)^n}  \right)^{\frac{1}{2}} e^{-\frac{1}{2}\beta (q_i - a_i)K_{ij} (q_j - a_j)} \nonumber \\
	& \equiv & Z(\bl)^{-1}e^{-\beta H(\mathbf{q}; \bl)}. 
\end{eqnarray}

If the system starts in equilibrium with respect to the initial values of the protocol $\bl(0)$ and is then driven out of equilibrium by executing the protocol at finite speed, the distribution will remain of a Gaussian form due to the linear nature of the forces \cite{mazonka1999exactly, van2003stationary}.  The non-equilibrium distribution $p_{\bl^*}$ can thus be written
\begin{eqnarray}
	p_{\bl^*}(\mathbf{q}) &=& \left(\frac{\beta \text{det}(\tilde{\mathbf{K}})}{(2\pi)^n}  \right)^{\frac{1}{2}} e^{-\frac{1}{2}\beta (q_i - \tilde{a}_i)\tilde{K}_{ij} (q_j - \tilde{a}_j)} \nonumber \\
	& \equiv & \tilde{Z}(\bl^*)^{-1}e^{-\beta \tilde{H}(\mathbf{q}; \bl^*) }
\end{eqnarray}
where $\tilde{K}_{ij} = K_{ij}(\bl^*_K) + \delta K_{ij}$ and $\tilde{a}_i = a_i(\bl^*_a) + \delta a_i$ are a ``lagged'' coupling matrix and mean vector.  The non-equilibrium exponential weight $\tilde{H}(\mathbf{q};\bl)$ can be written as 
\begin{equation}
	\tilde{H}(\mathbf{q};\bl) = H(\mathbf{q};\bl) + \delta H(\mathbf{X};\bl)
\end{equation}
where
\begin{equation}
	H(\mathbf{q};\bl) = \frac{1}{2}\left((q_i - a_i)K_{ij}(q_j - a_j) \right)
\end{equation}
is the Hamiltonian and
\begin{equation}
	\delta H(\mathbf{X};\bl) = \frac{1}{2}\left((q_i - a_i - \delta a_i)\delta K_{ij}(q_j-a_j-\delta a_j) - (q_i-a_i)K_{ij}\delta a_j - \delta a_iK_{ij}(q_j - a_j) + \delta a_i K_{ij} \delta a_j \right)
\end{equation}
is a lagged quantity.

We are interested in the gradients $\partial_\bl H(\mathbf{q}, \bl)$ and $\partial_\bl \tilde{H}(\mathbf{q}, \bl)$ which enter in to $\Delta^\text{eq}, \ \Delta^\text{app}$ and $\Delta^\text{neq}$.  For an element $\lambda_\alpha \in \bl_a$, we have
\begin{equation}
	\frac{\partial H }{\partial \lambda_\alpha} = \frac{\partial a_i }{\partial \lambda_\alpha}\frac{\partial H }{\partial a_i} = -\frac{\partial a_i }{\partial \lambda_\alpha} \left( K_{ij}(q_j - a_j)\right)
\end{equation}
and 
\begin{equation}
	\frac{\partial \tilde{H} }{\partial \lambda_\alpha} = \frac{\partial a_i }{\partial \lambda_\alpha}\frac{\partial \tilde{H} }{\partial a_i} = \frac{\partial a_i }{\partial \lambda_\alpha} \left(\frac{\partial H }{\partial a_i} + \delta\frac{\partial H }{\partial a_i}  \right) 
\end{equation}
where
\begin{equation}
	\delta\frac{\partial H }{\partial a_i} \equiv -\delta K_{ij}(q_j - a_j- \delta a_j) + K_{ij}\delta a_j.
\end{equation}
Similarly, for $\lambda_\alpha \in \bl_K$ we have
\begin{equation}
	\frac{\partial H }{\partial \lambda_\alpha} = \frac{\partial K_{ij} }{\partial \lambda_\alpha}\frac{\partial H }{\partial K_{ij}} = \frac{\partial K_{ij} }{\partial \lambda_\alpha} \left( \frac{1}{2}(q_i - a_i)(q_j - a_j)\right)
\end{equation}
and 
\begin{equation}
	\frac{\partial \tilde{H} }{\partial \lambda_\alpha} =\frac{\partial K_{ij} }{\partial \lambda_\alpha}\frac{\partial \tilde{H} }{\partial K_{ij}} =\frac{\partial K_{ij} }{\partial \lambda_\alpha} \left(\frac{\partial H }{\partial K_{ij}} + \delta\frac{\partial H }{\partial K_{ij}}  \right) 
\end{equation}
where 
\begin{equation}
	\delta\frac{\partial H }{\partial K_{ij}} \equiv -\frac{1}{2}\left((q_i - a_i)\delta a_j + \delta a_i(q_j - a_j) - \delta a_i \delta a_j\right).
\end{equation}
We next use these expressions to evaluate the gradients $\Delta^\text{eq}$, $\Delta^\text{neq}$, and $\Delta^\text{app}$.  For $\lambda_\alpha \in \bl_a$ have
\begin{eqnarray}
	\Delta ^\text{eq} &=& \llangle \beta \frac{\partial H(\mathbf{q}; \bl^n(t))}{\partial \lambda_\alpha} \rrangle_{p^\text{eq}_{\bl^*(t)}} - \llangle \beta \frac{\partial H(\mathbf{q}; \bl^n(t))}{\partial \lambda_\alpha} \rrangle_{p^\text{eq}_{\bl^n(t)}} \nonumber \\
	&=& -\beta \frac{\partial a_i^{n,t}}{\partial \lambda_\alpha}K_{ij}^{n,t}\left( \left(\llangle q_j \rrangle_{p^\text{eq}_{\bl^*(t)}} -a_j^{n,t}\right) - \left(\llangle q_j \rrangle_{p^\text{eq}_{\bl^n(t)}} -a_j^{n,t}\right) \right) \nonumber \\ 
	&=& -\beta \frac{\partial a_i^{n,t}}{\partial \lambda_\alpha}K_{ij}^{n,t}\left(a_j^{*,t} - a_j^{n,t} \right)
\end{eqnarray}
where we have introduced the shorthand notation $K_{ij}^{n,t} \equiv K_{ij}(\bl_K^n(t))$,  $a_i^{n,t} \equiv a_i(\bl_a^n(t))$, and $ \frac{\partial a_i^{n,t}}{\partial \lambda_\alpha} = \partial_{\lambda_\alpha} a_i|_{a_i = a_i^{n,t}}$.  We have also used the fact that $\llangle q_i \rrangle_{p^\text{eq}_{\bl^*(t)}} = a_i^{*,t}$ and $\llangle q_i \rrangle_{p^\text{eq}_{\bl^n(t)}} = a_i^{n,t}$.  We next have
\begin{eqnarray}
	\Delta ^\text{app} &=& \llangle \beta \frac{\partial H(\mathbf{q}; \bl^n(t))}{\partial \lambda_\alpha} \rrangle_{p_{\bl^*(t)}} - \llangle \beta \frac{\partial H(\mathbf{q}; \bl^n(t))}{\partial \lambda_\alpha} \rrangle_{p_{\bl^n(t)}} \nonumber \\
	&=& -\beta \frac{\partial a_i^{n,t}}{\partial \lambda_\alpha}K_{ij}^{n,t}\left( \left(\llangle q_j \rrangle_{p_{\bl^*(t)}} -a_j^{n,t}\right) - \left(\llangle q_j \rrangle_{p_{\bl^n(t)}} -a_j^{n,t}\right) \right) \nonumber \\ 
	&=& -\beta \frac{\partial a_i^{n,t}}{\partial \lambda_\alpha}K_{ij}^{n,t}\left(\left( a_j^{*,t} + \delta a_j^{*,t} -a_j^{n,t}\right) - \left(a_j^{n,t}+\delta a_j^{n,t} -a_j^{n,t}\right)\right) \nonumber \\
	&=& \Delta ^\text{eq}  -\beta \frac{\partial a_i^{n,t}}{\partial \lambda_\alpha}K_{ij}^{n,t}\left(\delta a_j^{*,t} - \delta a_j^{n,t} \right).
\end{eqnarray}
Here, we used the fact that $\llangle q_i \rrangle_{p_{\bl^*(t)}} = a_i^{*,t} + \delta a_i^{*,t}$ and $\llangle q_i \rrangle_{p_{\bl^n(t)}} = a_i^{n,t} + \delta a_i^{n,t}$.  Finally we have
\begin{eqnarray}
	\Delta ^\text{neq} &=& \llangle \beta \frac{\partial \tilde{H}(\mathbf{q}; \bl^n(t))}{\partial \lambda_\alpha} \rrangle_{p_{\bl^*(t)}} - \llangle \beta \frac{\partial \tilde{H}(\mathbf{q}; \bl^n(t))}{\partial \lambda_\alpha} \rrangle_{p_{\bl^n(t)}} \nonumber \\
	&=& \Delta ^\text{app} - \beta \frac{\partial a_i}{\partial \lambda_\alpha} \left(\delta K_{ij}^{n,t}\left(\left( \llangle q_j \rrangle_{p_{\bl^*(t)}} -a_j^{n,t} -\delta a_j^{n,t} \right) - \left( \llangle q_j \rrangle_{p_{\bl^n(t)}} -a_j^{n,t} -\delta a_j^{n,t} \right)\right) + K_{ij}^{n,t} \left(\delta a_{j}^{n,t} - \delta a_j^{n,t} \right)\right) \nonumber \\
	&=& \Delta ^\text{app} - \beta \frac{\partial a_i}{\partial \lambda_\alpha} \left(\delta K_{ij}^{n,t}\left(a_j^{*,t} + \delta a_j^{*,t} -  a_j^{n,t} -\delta a_j^{n,t}\right) \right).
\end{eqnarray}
We see that if there is no lag in the effective coupling matrix $\delta \mathbf{K}^{n,t}$, then $\Delta^\text{neq} = \Delta^\text{app}$ for $\lambda_\alpha \in \bl_a$.

For $\lambda_\alpha \in \bl_K$, we have 
\begin{eqnarray}
	\Delta ^\text{eq} &=& \beta \frac{1}{2}\frac{\partial K^{n,t}_{ij}}{\partial \lambda_\alpha}\left(\llangle(q_i - a^{n,t}_i)(q_j-a^{n,t}_j)\rrangle_{p^\text{eq}_{\bl^*(t)}} - \llangle(q_i - a^{n,t}_i)(q_j-a^{n,t}_j)\rrangle_{p^\text{eq}_{\bl^n(t)}} \right)  \nonumber \\
	&=& \beta \frac{1}{2}\frac{\partial K^{n,t}_{ij}}{\partial \lambda_\alpha}\bigg( \left(\llangle q_iq_j\rrangle_{p^\text{eq}_{\bl^*(t)}} -a_i^{n,t}\llangle q_j\rrangle_{p^\text{eq}_{\bl^*(t)}} - a_j^{n,t}\llangle q_i\rrangle_{p^\text{eq}_{\bl^*(t)}} + a_i^{n,t} a_j^{n,t}\right) \nonumber \\
	&& \ \ \ \ \ \ \ \ \ \ \ \ \ - \left(\llangle q_iq_j\rrangle_{p^\text{eq}_{\bl^n(t)}} -a_i^{n,t}\llangle q_j\rrangle_{p^\text{eq}_{\bl^n(t)}} - a_j^{n,t}\llangle q_i\rrangle_{p^\text{eq}_{\bl^n(t)}} + a_i^{n,t} a_j^{n,t}\right) \bigg) \nonumber \\
	&=& \beta \frac{1}{2}\frac{\partial K^{n,t}_{ij}}{\partial \lambda_\alpha}\left( 
	\llangle q_iq_j\rrangle_{p^\text{eq}_{\bl^*(t)}}  - \llangle q_iq_j\rrangle_{p^\text{eq}_{\bl^n(t)}} - (a_i^{n,t}a_j^{*,t} + a_j^{n,t}a_i^{*,t}) + 2 a_i^{n,t} a_j^{n,t} \right).
\end{eqnarray}
An alternative expression for this gradient can be found using definition of the correlation matrix $\mathbf{C}$
\begin{equation}
	C_{ij} \equiv \llangle(q_i - a_i)(q_j - a_j)\rrangle = \beta^{-1}(K^{-1})_{ij} 
\end{equation} 
for a multidimensional Gaussian parameterized by mean $\mathbf{a}$ and stiffness $\mathbf{K}$.  One can rewrite this as
\begin{equation}
	\llangle(q_i - b_i)(q_j - b_j)\rrangle = C_{ij} + (b_i - a_i)(b_j - a_j)
\end{equation}
for arbitrary $b_i$ and $b_j$.  With this can express $\Delta^\text{eq}$ as 
\begin{eqnarray}
	\Delta ^\text{eq} &=&  \beta \frac{1}{2}\frac{\partial K^{n,t}_{ij}}{\partial \lambda_\alpha}\left(C_{ij}^{*,t} - C_{ij}^{n,t} + (a_i^{n,t} - a_i^{*,t})(a_j^{n,t} - a_j^{*,t})  \right).
\end{eqnarray}
We next have
\begin{eqnarray}
	\Delta ^\text{app} &=&  \beta \frac{1}{2}\frac{\partial K^{n,t}_{ij}}{\partial \lambda_\alpha}\left(\llangle(q_i - a^{n,t}_i)(q_j-a^{n,t}_j)\rrangle_{p_{\bl^*(t)}} - \llangle(q_i - a^{n,t}_i)(q_j-a^{n,t}_j)\rrangle_{p_{\bl^n(t)}} \right) \nonumber \\
	&=& \beta \frac{1}{2}\frac{\partial K^{n,t}_{ij}}{\partial \lambda_\alpha}\left( \tilde{C}_{ij}^{*,t} + (a_i^{n,t}  - a_i^{*,t} - \delta a_i^{*,t})(a_j^{n,t} - a_j^{*,t} - \delta a_j^{*,t}) - \tilde{C}_{ij}^{n,t} - \delta a_i^{n,t} \delta a_j^{n,t} \right) \nonumber \\
	&=& \Delta^\text{eq} + \beta \frac{1}{2}\frac{\partial K^{n,t}_{ij}}{\partial \lambda_\alpha}\left( \delta C_{ij}^{*,t} - \delta C_{ij}^{n,t} - (a_i^{n,t} - a_i^{*,t})\delta a_j^{*,t} -  \delta a_i^{*,t}(a_j^{n,t} - a_j^{*,t}) \right)
\end{eqnarray}
where we have introduced $\tilde{C}_{ij} \equiv \beta^{-1}(\tilde{K}^{-1})_{ij} = C_{ij} + \delta C_{ij}$, with $\delta C_{ij}$ being the lag induced in the correlation matrix $\mathbf{C}$.  We note that the relation between $\delta K_{ij}$ and $\delta C_{ij}$ can be obtained using the Woodbury matrix identity (assuming that all matrices are invertible as necessary) 
\begin{eqnarray}
	\left(\mathbf{K} + \delta \mathbf{K} \right)^{-1} &=& \mathbf{K}^{-1} - \mathbf{K}^{-1}\left(\mathbf{K}^{-1} + (\delta \mathbf{K})^{-1}  \right)^{-1}\mathbf{K}^{-1} \nonumber \\
	&=& \beta \mathbf{C} - \beta^2 \mathbf{C} \left(\beta\mathbf{C} + (\delta \mathbf{K})^{-1} \right)^{-1}\mathbf{C} \nonumber \\
	&\equiv& \beta(\mathbf{C} + \delta{\mathbf{C}}).  
\end{eqnarray}
Assuming that the lag $\delta \mathbf{K}$ is small compared to $\mathbf{K}$, we can expand $\left(\beta \mathbf{C} + (\delta \mathbf{K})^{-1}  \right)^{-1} \approx \delta \mathbf{K}$ and obtain 
\begin{equation}
	\delta \mathbf{C} = - \beta \mathbf{C} (\delta \mathbf{K}) \mathbf{C}.
\end{equation}
Finally, we have
\begin{eqnarray}
	\Delta ^\text{neq} &=&  \Delta^\text{app} - \beta \frac{1}{2}\frac{\partial K^{n,t}_{ij}}{\partial \lambda_\alpha}\bigg(\llangle(q_i - a_i^{n,t})\delta a_j^{n,t} + \delta a_i^{n,t}(q_j - a_j^{n,t}) - \delta a_i^{n,t} \delta a_j^{n,t} \rrangle_{p_{\bl^*(t)}} \nonumber \\
	&& \ \ \ \ \ \ \ \ \ \ \ \ \ \ \ \ \ \ \ \ \ \ \ - \llangle(q_i - a_i^{n,t})\delta a_j^{n,t} + \delta a_i^{n,t}(q_j - a_j^{n,t}) - \delta a_i^{n,t} \delta a_j^{n,t} \rrangle_{p_{\bl^n(t)}} \bigg) \nonumber \\
	&=&  \Delta^\text{app} - \beta \frac{1}{2}\frac{\partial K^{n,t}_{ij}}{\partial \lambda_\alpha} \left((a_i^{*,t} -a_i^{n,t})\delta a_j^{n,t} + \delta a_i^{n,t}(a_j^{*,t} - a_j^{n,t})  +\delta a_i^{*,t}\delta a_j^{n,t} + \delta a_i^{n,t}\delta a_j^{*,t} - 2 \delta a_i^{n,t} \delta a_j^{n,t} \right).
\end{eqnarray}
We see that if there is no lag in the effective mean $\delta \mathbf{a}^{n,t}$, then $\Delta^\text{neq} = \Delta^\text{app}$ for $\lambda_\alpha \in \bl_K$.

\subsection{Fourier analysis of Helfrich membrane dynamics}
As described in the main text, the overdamped relaxational dynamics of the height field $q(r,t)$ of a Helfrich membrane are
\begin{equation}
	\partial_t q(r,t) = - \mu \frac{\delta H[q(r,t]}{\delta q(r,t)}= \mu \left(\left(\sigma + \frac{\kappa \lambda^2}{2}\right)\partial_r^2 q + \partial_r q \partial_r \lambda - \kappa\left(\partial_r^4 q - \partial_r^2 \lambda \right) \right). \label{eqqdynSI}
\end{equation}
We expand $q(r,t)$ in Fourier modes as
\begin{equation}
	q(r,t) = \sum_{m=-\infty}^{\infty} q_m(t) e^{imkr}
\end{equation}
where $k = 2\pi / L$.  Similarly, the protocol $\lambda(r,t)$ is expanded as 
\begin{equation}
	\lambda(r,t) = \sum_{m=-\infty}^{\infty} \lambda_m(t) e^{imkr}.
\end{equation}
For small values of $\lambda(r,t)$ and $q(r,t)$, we neglect the nonlinear terms in Equation \ref{eqqdynSI} and write the equation of motion for the amplitude of the $m^\text{th}$ mode as
\begin{eqnarray}
	\partial_t q_m(t) &=& -\mu m^2 k^2 \left((\sigma  + m^2 k^2 \kappa) q_m(t) + \kappa \lambda_m(t)  \right) \nonumber \\
	&=& - \mu K_m(q_m(t) - a_m(t)) \label{eqHelfFour}
\end{eqnarray}
where 
\begin{equation}
	K_m \equiv m^2k^2(\sigma + m^2 k^2\kappa)
\end{equation}
and 
\begin{equation}
	a_m(t) \equiv -\frac{\kappa }{\sigma + m^2 k^2\kappa}\lambda_m(t).
\end{equation}
Comparing Equations \ref{eqLinDyn} and \ref{eqHelfFour} see that the $m^\text{th}$ modes acts like a particle in a harmonic trap whose position $a_m(t)$ but not stiffness $K_m$ is altered as a function of time.  The coupling matrix $K_{ij}$ in this case is diagonalized due to the neglected non-linear terms in the membrane dynamics, and we have also neglected the noise; these extensions to the theory of trainable membranes could be treated in future work.

\section{Active nematic defect control}\label{secSInematic}

\subsection{Equations of motion} 
Nematic systems are described by a symmetric and traceless tensor
\begin{equation}
	\mathbf{Q} = q(\hat{\mathbf{n}}\hat{\mathbf{n}} - \frac{1}{d}\mathbf{I})
\end{equation}
where $\hat{\mathbf{n}}$ is a unit director, $q$ quantifies the polarization of the nematic, $d$ is the system dimensionality, and $\mathbf{I}$ is the identity tensor.  This order parameter $\mathbf{Q}$ couples to a flow field $\mathbf{v}$ and relaxes along the gradient of a free energy function.  We study a simplified version of these physics, taking the overdamped limit and the limit of high substrate friction \cite{shankar2022topological}.  We have the equations of motion
\begin{align}
	\partial_t Q_{ij} =& S_{ij}(\mathbf{v}) + \Gamma_H H_{ij} \label{eqQdyn} \\
	v_i =& \gamma_{v}^{-1} \partial_k\left(\sigma^a_{ik}(\mathbf{Q}) + \sigma^E_{ik}(\mathbf{Q}) \right). \label{eqv}
\end{align}
Here, $H_{ij}$ is the symmetric traceless part of $-\frac{\delta F}{\delta Q_{ij}}$ with $F$ the free energy, $S_{ij}$ is a flow-coupling term, and $\gamma_v$ is a friction coefficient.  In these overdamped dynamics, $\mathbf{v}$ is given instantaneously in terms of $\mathbf{Q}$ so that Equation \ref{eqQdyn} is closed in $\mathbf{Q}$.  The active and Ericksen stress tensors are \cite{zhang2021spatiotemporal}
\begin{align}
	\sigma^a_{ij} =& -\alpha Q_{ij} \\
	\sigma^E_{ij} =& f\delta_{ij} -\lambda H_{ik}\left(Q_{kj}+\frac{1}{3}\delta_{kj} \right) - \lambda\left(Q_{ik}+\frac{1}{3}\delta_{ik} \right)H_{kj} \nonumber \\
	&+2\lambda \left(Q_{ij}+\frac{1}{3}\delta_{ij} \right)H_{kl}Q_{kl} - \partial_j Q_{kl}\frac{\delta F}{\delta \partial_i Q_{kl}} \nonumber \\ &+ Q_{ik}H_{kj} - H_{ik}Q_{kj}.
\end{align}
The Landau-de Gennes free energy is 
\begin{equation}
	F = \int d \mathbf{r} f(\mathbf{r})
\end{equation}
where 
\begin{align}
	f = \frac{A_0}{2}\left(1-\frac{U}{3} \right)\text{Tr}\left(\mathbf{Q}^2\right) -\frac{A_0 U}{3}\text{Tr}\left(\mathbf{Q}^3\right) \nonumber \\
	+ \frac{A_0 U}{4}\text{Tr}\left(\mathbf{Q}^2\right)^2 + \frac{L}{2}\left(\partial_kQ_{lm}\right)^2.
\end{align}
Finally, the flow coupling term is 
\begin{align}
	S_{ij}(\mathbf{v}) =& -v_k \partial_k  Q_{ij} + \Phi_{ik} Q^+_{kj} +Q^+_{ik}\Phi_{kj} \nonumber - 2\lambda Q^+_{ij} \left(Q_{kl}\partial_k v_l\right),
\end{align}
where
\begin{equation}
	Q^+_{ij} = Q_{ij} + \frac{1}{3}\delta_{ij},
\end{equation}
\begin{equation}
	\Psi_{ij} = \frac{1}{2}\left(\partial_i v_j + \partial_j v_i \right),
\end{equation}
\begin{equation}
	\Omega_{ij} = \frac{1}{2}\left(\partial_i v_j - \partial_j v_i \right),
\end{equation}
and
\begin{equation}
	\Phi_{ij} = \xi \Psi_{ij} - \Omega_{ij}.
\end{equation}
In these equations, $\xi$, $A_0$, $U$, $\Gamma_H$, $\gamma_v$ are parameters whose meaning is described in Ref.\ \citenum{zhang2021spatiotemporal}.  We set $\xi = 0.7$, $A_0 = 0.1$, $U = 3.5$, $\Gamma_H = 1.5$, and $\gamma_v = 10$ (all in simulation units).  

\subsection{Stabilization} 
Although the update in Equation \ref{eqnematicupdate} of the main text provides sufficient information to reconstruct a target defect trajectory, it can also lead to unwanted behavior if not stabilized.  In particular, large activity and nematic gradients can cause nucleation of new defects, which we treat as a terminal condition in which the learning dynamics have failed.  To prevent this, we stabilize the learning dynamics in three ways:
\begin{itemize}
	\item We encourage updates to $\alpha(\mathbf{r},t)$ only in the near vicinity of the defect through an eligibility trace-like field \cite{sutton2018reinforcement} $z(\mathbf{r},t)$, which evolves during learning as
	\begin{equation}
		z^{n+1}(\mathbf{r},t) \leftarrow \lambda_z z^{n+1}(\mathbf{r},t) + f^n(\mathbf{r},t) / \bar{f}
	\end{equation}
	where $\lambda_z \leq 1$ is a decay factor and $\bar{f} = 0.01$ is scale factor for the free energy density.  This eligibility trace multiplies the term $(f^*(\mathbf{r},t) -  f^n(\mathbf{r},t))$ in Equation \ref{eqnematicupdate} of the main text.  Because defects correspond to persistent peaks in $f(\mathbf{r},t)$, the eligibility $z$ will continually be supported near defects and will otherwise decay to zero at a rate $\sim 1/\lambda_z$.  We set $\lambda_z = 0.75$ for the results in the main text.  
	\item We only allow learning when the trial defect position is within a certain spatial window of the target defect position.  In principle the ``zippering'' mechanism discussed in SI Section \ref{sectransmat} guarantees convergence for the whole trajectory, because eventually the first part of the trajectory will be learned, after which the second part has the correct starting configuration and can in turn converge, and so on.  However, during early learning iterations the activity updates for later parts of the trajectory (when the defect is not close to its target) can lead to unstable spots which can nucleate new defects.  On the other hand, when the learning has successfully brought the defect position close to its target, the activity fields can still evolve in general because the two free energy profiles do not match exactly.  This effect can also destabilize learning.  We thus turn off learning when the defect position is too far ($>5$ lattice units) or too close ($< \sqrt{2}$ lattice units) to the target defect position.  
	\item We clamp the local absolute value of activity at $|\alpha(\mathbf{r},t)|_\text{max} = 12$, and we turn off learning when the total absolute activity in the system has passed $\int d\mathbf{r}|\alpha(\mathbf{r},t)| = 150$. 
\end{itemize}
It remains to explore whether all of these stabilization methods are strictly necessary or whether they could be further refined; we simply found that this combination seems to work well for our test cases.

\subsection{Explanation of the update rule}
Here we draw on recent theoretical work \cite{shankar2022spatiotemporal} to justify the efficacy of Equation \ref{eqnematicupdate} in the main text.  The authors of Ref.\ \citenum{shankar2022spatiotemporal} have shown that the approximate velocity of $-1/2$ defect due to a spatially inhomogenous activity field is
\begin{equation}
	v^-_i = a \Theta_{ijk} \partial_j \partial_k \alpha. \label{eqminusvel}
\end{equation}
Here,
\begin{equation}
	\Theta_{ijk} = \hat{t}_i\hat{t}_j\hat{t}_k - \frac{1}{4}\left(\delta_{ij}\hat{t}_k +\delta_{kj}\hat{t}_i+\delta_{ik}\hat{t}_j\right)
\end{equation}
is a rank-three tensor describing the orientation of a defect which has one of its three legs directed along ${\hat{\mathbf{t}}=(\cos(\theta),\sin(\theta))}$.  The orientation tensor is invariant under $\theta \leftarrow \theta + m 2\pi/3$ for any integer $m$, and it is equal to zero under contraction of any two of its indices.  Denoting $g_{ij} = \partial_i \partial_j \alpha$, we can express Equation \ref{eqminusvel} as
\begin{equation}
	\mathbf{v}^- =  \mathbf{M} \cdot \mathbf{c} \label{eqminusvelsimp}
\end{equation}
where
\begin{equation}
	\mathbf{M} = \frac{1}{4}\begin{pmatrix}
		g_{xx} - g_{yy} & g_{xy} + g_{yx} \\
		-(g_{xy} + g_{yx}) & g_{xx} - g_{yy}
	\end{pmatrix}
\end{equation}
and $\mathbf{c} = (\cos(3\theta), \sin(3\theta))$.

Referring to Figure \ref{NematicResultsComposite}E of the main text, if we slightly displace a defect, whose free energy profile is $f(\mathbf{r})$, along a vector $\mathbf{d}$, it will have an approximate free energy profile 
\begin{equation}
	f^*(\mathbf{r}) = f(\mathbf{r} - \mathbf{d}) \approx f(\mathbf{r}) - d_i\partial_i f(\mathbf{r}) +\frac{1}{2}d_id_j\partial_i\partial_j f(\mathbf{r}).
\end{equation}
Considering an activity update
\begin{equation}
	\alpha(\mathbf{r},t) \leftarrow \alpha(\mathbf{r},t) - \eta (f^*(\mathbf{r},t) - f(\mathbf{r},t))
\end{equation}
we can write for the first iteration (following the initial guess $\alpha(\mathbf{r},t) = 0$)
\begin{equation}
	\alpha(\mathbf{r},t) =  \eta \left(d_i\partial_i f(\mathbf{r}) - \frac{1}{2}d_id_j\partial_i\partial_j f(\mathbf{r})\right ).
\end{equation}
Evaluating $g_{ij}$, we have (setting $\eta = 1$ for simplicity)
\begin{equation}
	g_{ij} = d_k \partial_i \partial_j \partial_k f(\mathbf{r}) - \frac{1}{2} d_k d_l  \partial_l \partial_k \partial_i \partial_j f(\mathbf{r}).
\end{equation}
We will evaluate this expression at the location of the defect $\mathbf{r} = \mathbf{0}$ where the free energy profile is maximal and assume that the free energy profile is approximately isotropic \cite{zhou2017fine}.  The third order derivatives of $f(\mathbf{r})$ consequently vanish, so that
\begin{equation}
	g_{ij}(\mathbf{0}) = - \frac{1}{2} d_k d_l \partial_l \partial_k \partial_i \partial_j f(\mathbf{0}).
\end{equation}
The only terms which can contribute to this sum are those having even numbers of derivatives with respect to $x$ and $y$ (i.e. $\partial_{x}^3\partial_y f(\mathbf{0}) = 0$ but $\partial_{x}^2\partial_y^2 f(\mathbf{0}) \neq 0$).  For an isotropic function $f(\mathbf{r}) = f(r)$ one can show that $\partial_x^4 f(\mathbf{0}) = \partial_y^4 f(\mathbf{0}) = 3 \partial_x^2 \partial_y^2 f(\mathbf{0})$.  Denoting $\partial_x^2 \partial_y^2 f(\mathbf{0}) \equiv f'' > 0$ and $\mathbf{d} = d(\cos(\phi), \sin(\phi))$, we evaluate Equation \ref{eqminusvelsimp} 
\begin{equation}
	\mathbf{v}^- = \frac{1}{4}d^2f''\begin{pmatrix}
		-\cos(2 \phi - 3\theta) \\
		\sin(2\phi - 3\theta) 
	\end{pmatrix}.
\end{equation}
Finally, we are interested in the overlap of this defect velocity with the displacement vector $\mathbf{d}$.  We thus compute
\begin{equation}
	\mathbf{d} \cdot \mathbf{v}^- = -\frac{1}{4}d^3f''\cos(3\psi).
\end{equation}
where $\psi = \phi - \theta$ is the relative angle between the defect orientation and $\mathbf{d}$.  As $f''$ is negative, we have that the overlap is proportional to $\cos(3\psi)$.

\section{Connection to ideas in reinforcement learning}\label{secSIRL}
Reinforcement learning (RL) refers to a class of methods in which an agent learns through trial and error how to exert actions on its environment in order to maximize a user-defined reward \cite{sutton2018reinforcement}. Here, we briefly outline the conceptual similarities and differences between techniques which are utilized in standard RL algorithms and those discussed in this paper.     

\subsection{States, actions, rewards, and value functions}
The standard setting of RL comprises a set of states, a learner which can exert actions on the system, a dynamical rule for the system which maps a state and an action into a new state (either deterministically or stochastically), and a reward function.  The reward is determined by the user and in principle is completely arbitrary, although in practice a judicious specification of the reward structure is crucial to a successful RL implementation.  Rewards are often chosen to satisfy some criterion of the system, such as to balance a pole against gravity in the famous cart-pole task.  

In this paper we consider deterministic dynamics, either operating on probability distributions or on individual average configurations of the system.  Actions correspond to time-dependent control fields which drive the system through non-equilibrium trajectories.  In the language of RL, our ``reward functions'' pose an open-loop, inverse problem in which we want to minimize the difference between the current system trajectory and a given target trajectory.  We use both KL divergences and quadratic cost functions in this paper to quantify this difference.  

\subsection{Learning the value function}
The agent chooses an action according to its current policy, which can be either a stochastic or deterministic function of the current state.  
RL algorithms optimize the expected reward during training by incrementally improving the policy using new experiences.  A standard algorithm for updating a deterministic policy was introduced in Ref.\ \citenum{silver2014deterministic}.  Specifically, for a policy $\mu_{\boldsymbol{\theta}}(s)$, which is a function that maps a state $s$ into an action $a$ and has learnable parameters $\boldsymbol{\theta}$, updates happen according to 
\begin{equation}
	\boldsymbol{\theta}^{n+1} \leftarrow  \boldsymbol{\theta}^{n} + \eta \nabla_{\boldsymbol{\theta}}Q(s_t, a_t)|_{\boldsymbol{\theta} = \boldsymbol{\theta}^n}. \label{eqSIRLupdate}
\end{equation}
Here, the current state is $s_t$ and the agent has taken action $a_t$.  The function $Q(s, a)$ is called the value function, which reflects the agent's current estimate of the cumulative reward that it will receive if it takes action $a$ while in state $s$.  It depends implicitly on the parameters $\boldsymbol{\theta}$ via the action selection $a = \mu_{\boldsymbol{\theta}}(s)$, and as a result the gradient $\nabla_{\boldsymbol{\theta}}Q(s_t, a_t)$ can be expanded using the chain rule.  

Equation \ref{eqSIRLupdate} is a simple learning rule which increments $\boldsymbol{\theta}$ so that it will increase expected reward (as estimated through its learned value function).  Clearly, the success of RL will depend on how well $Q(s,a)$ can be learned by the agent.  A common approach is to represent $Q(s,a)$ by a parameterized function $Q^\mathbf{w}(s,a)$ and to increment its parameters $\mathbf{w}$, in addition to $\boldsymbol{\theta}$, according to a separate update rule.  A standard rule for $\mathbf{w}$ is the semi-gradient temporal difference (TD) scheme \cite{sutton2018reinforcement, silver2014deterministic}
\begin{equation}
	\mathbf{w}^{n+1} \leftarrow \mathbf{w}^n + \eta_w\left(r_t + \gamma Q^\mathbf{w}(s_{t+1}, a_{t+1}) - Q^\mathbf{w}(s_{t}, a_{t})\right) \nabla_\mathbf{w}Q^\mathbf{w}(s_t,a_t)|_{\mathbf{w} = \mathbf{w}^n} \label{eqSIwupdate}
\end{equation}
where $\gamma$ is a discount factor, $\eta_w$ is a learning rate, and $r_t$ is the reward received at time $t$.  The TD approximation in this update rule results from truncating the Bellman equation one timestep into the future (see SI Section \ref{sectemploc}), and the approximation leading to semi-gradient methods results from neglecting the dependence of $Q^\mathbf{w}(s_{t+1}, a_{t+1})$ on the current value of $\mathbf{w}$.  Using this update rule works well in practice, but a practical downside is that it requires separate storage for the parameters $\mathbf{w}$.

In this context, our methods can be loosely viewed as bypassing the need to separately learn the value function $Q(s,a)$, by replacing $\nabla_{\boldsymbol{\theta}}Q(s_t, a_t)$ in Equation \ref{eqSIRLupdate} with a prescribed, physically approximated error signal $\sim\Delta^\text{app}$.  In Equation \ref{eqSIwupdate}, approximations are used to update the current estimate of the value function, whereas in our approach approximations are used in writing down the gradient of the value function directly.  We do assume knowledge of the target trajectory in writing down $\Delta^\text{app}$, but this information is used in Equation \ref{eqSIwupdate} to determine the rewards $r_t$.  

\subsection{Exploiting temporal locality} \label{sectemploc}
Both our learning rules and those based on the TD method are temporally local, in that they consider at most a few timesteps and not the entire system trajectory.  As mentioned above, updates to the parameters $\mathbf{w}$ of a learned value function $Q^\mathbf{w}(s,a)$ exploit temporal locality through the TD error $\delta_t = r_t + \gamma Q^\mathbf{w}(s_{t+1}, a_{t+1}) - Q^\mathbf{w}(s_{t}, a_{t})$, which will be small if $Q^\mathbf{w}(s_{t}, a_{t}) \approx r_t + \gamma Q^\mathbf{w}(s_{t+1}, a_{t+1})$.  This condition is a statement of the Bellman equation, which roughly says that the cumulative expected reward starting at time $t$ is equal to the reward accrued at $t$ plus the discounted cumulative expected reward starting at time $t+1$.  Thus, a converged estimate for the value function will satisfy the Bellman equation and lead to small TD errors.  

The justification of temporally local update rules in our method is reminiscent of but slightly different from the Bellman equation.  In our inverse problem setup, a perfect gradient would include the effect of updating actions at time $t-1$ on the loss both at time $t$ and at all future times $t' \geq t$ (cf.\ Equation \ref{eqLossfulltime} above).  We avoid the latter contribution because it is non-local and involves backpropagation of errors through time.  This is justified because of the zippering mechanism of convergence, outlined in SI Section \ref{sectransmat}.  One can view this mechanism as endowing the inverse problem with so-called ``optimal substructure,'' in which a complex optimality problem (learning the whole trajectory) can be decomposed into a set of optimality sub-problems (learning individual time points or successive differences).  By a similar token, the Bellman equation gives optimal substructure to the problem of maximizing the cumulative expected reward, which justifies the TD method of RL.  Whereas in TD the optimal substructure allows updating the estimate of the value function using temporally local information, for us the optimal substructure allows following a physically prescribed value function gradient in a temporally local manner.

\bibliographystyle{unsrt}

\end{document}